%% file: main.tex
\UseRawInputEncoding
\documentclass[aps, prc, twocolumn, superscriptaddress, floatfix, nofootinbib]{revtex4-2}

\usepackage{graphicx}
\usepackage{longtable}
\usepackage{siunitx}
\usepackage{multirow} 

\begin{document}

\title{Core Breaking at Low Spin in $^{68}$Zn from Nuclear Resonance Fluorescence}

\author{S. R. Johnson}
\altaffiliation[Present address: ]
{Department of Physics, North Carolina State University, Raleigh, North Carolina 27695-8202, USA and Triangle Universities Nuclear Laboratory, Duke University, Durham, North Carolina 27708, USA}
\author{R. V. F. Janssens}
\affiliation{Department of Physics \& Astronomy, University of North Carolina at Chapel Hill, Chapel Hill, North Carolina 27599, USA}
\affiliation{Triangle Universities Nuclear Laboratory, Duke University, Durham, North Carolina 27708, USA}
\author{B. A. Brown}
\affiliation{Facility for Rare Isotope Beams, Michigan State University, East Lansing, Michigan 48824, USA}
\affiliation{Department of Physics \& Astronomy, Michigan State University, East Lansing, Michigan 48824, USA}

\author{A. D. Ayangeakaa}
\affiliation{Department of Physics \& Astronomy, University of North Carolina at Chapel Hill, Chapel Hill, North Carolina 27599, USA}
\affiliation{Triangle Universities Nuclear Laboratory, Duke University, Durham, North Carolina 27708, USA}

\author{S. S. Bhattacharjee}
\affiliation{TRIUMF, 4004 Wesbrook Mall, Vancouver, British Columbia, Canada V6T 2A3}

\author{E. Churchman}
\affiliation{Department of Physics \& Astronomy, University of North Carolina at Chapel Hill, Chapel Hill, North Carolina 27599, USA}
\affiliation{Triangle Universities Nuclear Laboratory, Duke University, Durham, North Carolina 27708, USA}

\author{S. W. Finch}
\affiliation{Department of Physics, Duke University, Durham, North Carolina 27708, USA}
\affiliation{Triangle Universities Nuclear Laboratory, Duke University, Durham, North Carolina 27708, USA} 

\author{U. Friman-Gayer}
\affiliation{Department of Physics, Duke University, Durham, North Carolina 27708, USA}
\affiliation{Triangle Universities Nuclear Laboratory, Duke University, Durham, North Carolina 27708, USA}

\author{S. Frye}
\author{M. Fulghieri}
\altaffiliation[Present address: ]
{Laboratory for Nuclear Science, Massachusetts Institute of Technology, Cambridge, Massachusetts, 02139, USA}
\author{D. Gribble}
\author{X. H.-K. James}
\affiliation{Department of Physics \& Astronomy, University of North Carolina at Chapel Hill, Chapel Hill, North Carolina 27599, USA}
\affiliation{Triangle Universities Nuclear Laboratory, Duke University, Durham, North Carolina 27708, USA}

\author{R. Longland}
\affiliation{Department of Physics, North Carolina State University, Raleigh, North Carolina 27695-8202, USA}
\affiliation{Triangle Universities Nuclear Laboratory, Duke University, Durham, North Carolina 27708, USA}

\author{C. Wegner}
\affiliation{Department of Physics \& Astronomy, University of North Carolina at Chapel Hill, Chapel Hill, North Carolina 27599, USA}
\affiliation{Triangle Universities Nuclear Laboratory, Duke University, Durham, North Carolina 27708, USA}

\date{\today}

\begin{abstract}
    Low-spin excited states in $^{68}$Zn have been studied at the High Intensity Gamma-Ray Source (HI$\gamma$S) from the ground state up to the particle emission threshold using the nuclear resonance fluorescence technique (NRF) and the newly developed Clover Array. Low-spin levels were excited by linearly-polarized, $2.90 – 9.79$~MeV photon beams. Spin-parity quantum numbers as well as associated $M1$ and $E1$ decay strengths were determined for a large fraction of the 158 states observed. In addition, long-duration coincidence measurements at 9.46 and 9.79~MeV enabled the investigation of the level scheme near the ground state. The results have been interpreted with shell-model calculations using two different model spaces and several effective interactions often used to describe nuclei in this mass region. While the structure near the ground state can be understood in terms of excitations involving solely valence nucleons, core breaking is required to account for the evolution of the total $M1$ strength at excitation energies above $\sim5$~MeV.
\end{abstract}

\maketitle

\section{Introduction}
The Ni isotopic chain at the $Z=28$ shell closure has been the subject of intense recent theoretical and experimental study. Unique impacts of the monopole tensor interaction have been demonstrated in this chain and in neighboring nuclei in the region. Early on, signatures of this interaction were seen mainly in nuclei far from stability \cite{Huck1985, Dinca2005, Janssens2002, Liddick2004, Liddick2004_2, Fornal2004, Prisciandaro2001, Burger2005, Al-Khalili2000, Otsuka2001, Karthika2021}, where extreme proton-to-neutron ratios create unexpected shell closures as a result of shell evolution \cite{Otsuka2005}. The more subtle effects of the interaction are highlighted in the presence of coexisting shapes in the Ni isotopes \cite{Prokop2015, Otsuka2016, Broda2012, Suchyta2014, Tsunoda2014, Leoni2017, Olaizola2017}.

Experimental properties of the excited- and ground-state $0^+$ configurations of the neutron-rich Ni isotopes could be described with Monte Carlo shell-model calculations (MCSM) that explicitly include the monopole tensor interaction \cite{Crider2016, Prokop2015, Leoni2017, Olaizola2017, Otsuka2016, Tsunoda2014}. Recently, a similar picture of shape coexistence was established for $^{64}$Ni, for the first time in a stable isotope. A variety of experimental probes were used, including a nuclear resonance fluorescence measurement in which all excited $0^+$ states were populated, thus characterizing states associated with spherical, oblate, and prolate minima \cite{Marginean2020,Little2022}.

The present work aims to lay the groundwork for a similar multi-pronged investigation of the Zn isotopic chain, beginning with nuclear resonance fluorescence (NRF) measurements. In these isotopes, two additional protons beyond the $Z=28$ shell closure can probe the shape-driving properties of the occupied orbitals. It is unclear whether the aforementioned coexisting shapes of the Ni isotopes persist with these additional protons. In $^{68}$Zn, the isotone of unstable, shape-isomeric $^{66}$Ni \cite{Leoni2017}, preliminary MCSM calculations also predict the presence at relatively low excitation energy of four $0^+$ states associated with different deformations. Some previous studies of $^{68}$Zn indicate that deformation is present at high spin \cite{McCutchan2012,Devlin1999} and a soft triaxial configuration has been proposed in Ref. \cite{Koizumi2004}  for the $0^+_2$ state at 1656~keV. To determine more completely the low-spin structure of $^{68}$Zn, experimental data is needed beyond what is currently available \cite{McCutchan2012}.

NRF is expected to be an effective probe of the low-energy level structure of $^{68}$Zn with long-duration measurements generating high-statistics $\gamma-\gamma$ coincidence datasets, as was the case for a recent study of $^{64}$Ni \cite{Marginean2020}. The coincidence approach utilizes the NRF technique somewhat indirectly as it relies on the average decay of high-energy, low-spin photoexcited states to populate the low-energy levels of interest. Furthermore, the low-spin photoresponse of the nucleus can itself provide structural insights. For example, a recent NRF study of $^{74}$Ge \cite{Johnson2023} tested the validity of two different shell-model effective interactions by comparing predictions of photonuclear cross section values for spin-1 states to those found experimentally. The study found good agreement between the experiment and theory up to $\sim5.6$~MeV. Extended NRF measurements of $^{68}$Zn present an opportunity to test the same theoretical approach, specifically in a low-spin domain, but spanning all possible excitation energies up to the particle-emission threshold.

Thus far, only a handful of excited states have been studied in $^{68}$Zn using the NRF technique, the most recent measurement dating back to 1983 \cite{Moreh1983}. In contrast, extensive NRF measurements were recently carried out on the $^{66}$Zn isotope \cite{Savran2022, Schwengner2021}, revealing 128 spin-1 states and establishing their excitation energy and parity as well as the associated $E1$ and $M1$ strength functions. Given this wealth of data, a high-statistics comparative NRF study of $^{68}$Zn will document both changes in structure and in $\gamma$-ray strength brought about by the addition of two neutrons. In the present work, these objectives are supported by the first use at the High Intensity Gamma-ray Source (HI$\gamma$S) of the Clover Array, a new multi-detector system providing enhanced capabilities over earlier NRF setups in terms of energy resolution, detection efficiency, and timing as well as in the ability to carry out high-statistics coincidence measurements.

This paper is divided into five sections. A description of the experiment and associated analysis techniques is given in Section II, which is followed by the presentation of the results (Section III). A discussion of the latter in comparison with data on neighboring nuclei and with shell-model calculations is provided in Section IV, and conclusions are drawn from the present work in Section V.

\section{Methods}

The photoexcitation study was carried out in two parts. Excited states were populated in a 1.8 g metallic $^{68}$Zn target, enriched to 98.5\%, using the High Intensity $\gamma$-Ray Source (HI$\gamma$S) at the Triangle Universities Nuclear Laboratory \cite{Weller2009}. In the first phase, the photon beam energy was varied in 34 steps from 2.90 to 9.79~MeV (referred to as the ``NRF scan" below), with a $\sim3\%$ FWHM energy spread, scanning the entire spin-1 excitation energy range of $^{68}$Zn below the particle-emission threshold at 9977~keV \cite{McCutchan2012}. Then, in a second step, dedicated long-duration coincidence measurements were performed at the two highest beam energies, 9.46 and 9.79~MeV. In both cases, the linearly-polarized photon beam was collimated to a 1.9~cm diameter spot with a post-collimation flux of $\sim10^7$ photons per second.

Fluorescent radiation emitted by the target was measured using the Clover Array, consisting of eight clover-type high-purity germanium detectors (HPGe) and 12 CeBr$_3$ scintillators \cite{Ayangeakaa2021}. Germanium detectors were positioned at the following locations: $(\theta, \phi) = (90^\circ,0^\circ), (90^\circ,90^\circ), (135^\circ,0^\circ), (125.26^\circ, 45^\circ)$ and reflection-symmetric positions, where $\theta$ is the polar angle with respect to the beam propagation direction and $\phi$ is the azimuthal angle with respect to the direction of beam polarization. Scintillators were placed at gaps between the clover detectors at a variety of azimuthal angles in the $\theta=90^\circ,135^\circ$ planes. Thin (1 mm) absorbers of copper and lead were placed on the front face of all detectors in the array to attenuate x-rays and low-energy $\gamma$ rays. A detailed description of the array as well as data analysis techniques can be found in Ref. \cite{Johnson2025}. 

High-resolution timing and energy information for each above-threshold signal from a scintillator or clover segment was digitized. Clover detector data was processed in addback mode; i.e., the energy deposited in every segment of a given clover was summed for each beam pulse. This approach increases the efficiency of the system and reduces the impact of electron/positron escapes and of scattering processes occurring in the detector material.

\subsection{NRF Scan}

For each beam energy in the NRF scan from 2.90 to 9.79~MeV, $\gamma$ rays in the clover spectra occurring within the energy range covered by the beam were identified as ground-state (or ``elastic") decays from excited states in $^{68}$Zn. Lower-energy peaks in the same data were associated with transitions to excited states, or ``inelastic" decay branches. The energy and angular distribution of every transition was determined using the clover spectra. A quadratic energy calibration was performed using $\gamma$ rays from $^{152}$Eu and $^{56}$Co radioactive sources for the low beam-energy data and known elastic transition lines in $^{68}$Zn (3346.09(20), 4992.0(10), and 5298.0(4)~keV \cite{McCutchan2012}), $^{56}$Fe (6926.0(20), 7211.5(20), and 7285.8(4)~keV \cite{Junde2011}), and $^{32}$S (8125.40(20)~keV \cite{Ouellet2011}). These latter two targets were placed in the beam during this experimental campaign for preliminary testing for future experiments. Efficiency calibrations utilized a \textsc{Geant4} simulation of the array \cite{FrimanPapst2022}, scaled to data from the same radioactive sources. Level energies reported in this work are recoil-corrected weighted averages of the $\gamma$-ray energies measured at each detector position. Covariance in the energy calibration cannot be neglected at high beam energies, and has been incorporated in the uncertainty of the reported results.

\begin{figure}
    \centering
         \includegraphics[width=\linewidth]{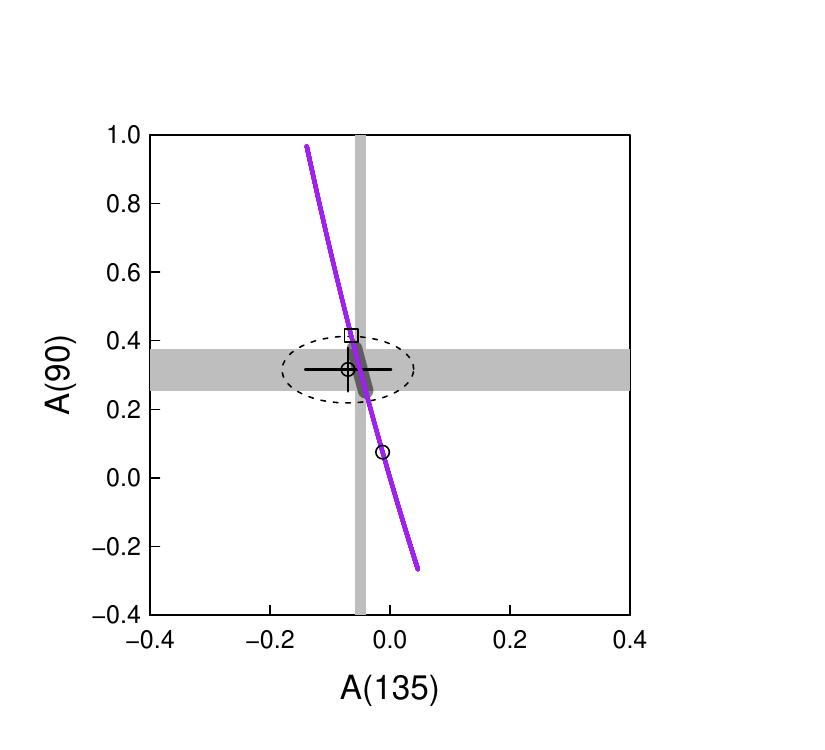}
    \caption{Example of a multipole mixing ratio determination for the transition from the NRF level at 6215 keV with $J^\pi=1^-$ to the $2^+_1$ state at 1077 keV. The purple curve represents all possible values of multipole mixing ratio $\delta$ in the phase space $A(135^\circ)$ vs $A(90^\circ)$. The location of $\delta=\pm\infty$ is given by the open square, and that of $\delta=0$ by the open circle. Experimental asymmetries are plotted in black, with the dashed line representing the associated error ellipse. One of two Bayesian best-fit values for $\delta$ is highlighted in dark gray, with projections onto either axis in light gray. See text for more details.}
    \label{fig:DeltaFit}
\end{figure}

\begin{figure}
    \centering
    \includegraphics[width=\linewidth]{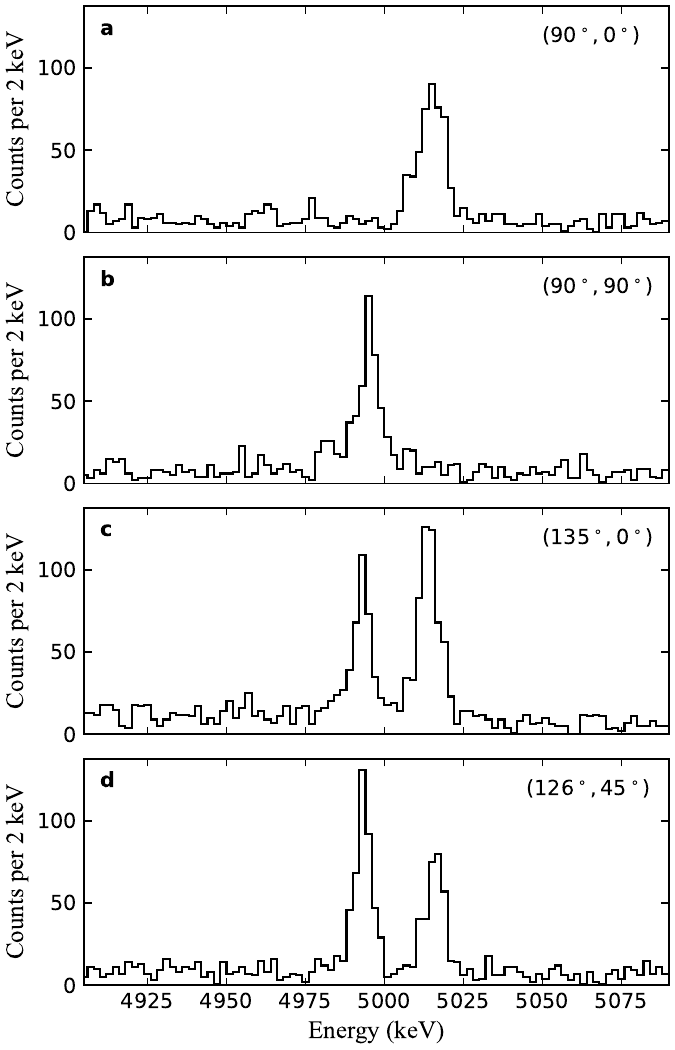}
    \caption{Elastic transitions from two NRF levels at 4993 and 5015~keV as measured by detectors at each position $(\theta,\phi)$: (a) ($90^\circ$,$0^\circ$), (b) ($90^\circ$,$90^\circ$), (c) ($135^\circ$,$0^\circ$), and (d) ($125.26^\circ$, $45^\circ$).}
    \label{fig:spectra_sample}
\end{figure}

Taking advantage of the $>99\%$ linear polarization of the beam, angular distribution data for elastic transitions was used to assign spin-parity values to the excited states using the experimental asymmetry
\begin{equation}
    \Sigma = \frac{N_\parallel-N_\perp}{N_\parallel+N_\perp}
\end{equation}
with $N_\parallel$ and $N_\perp$ being the efficiency-corrected counts for a $\gamma$ ray detected parallel and perpendicular to the beam polarization plane, respectively. This experimental quantity is compared to the asymmetry of the angular distribution function $W(\theta,\phi)$ for the decay of states with spin and parity values $J^\pi=1^\pm,2^\pm$, with Q-coefficient solid angle corrections calculated with Monte Carlo simulations. Asymmetry data for the inelastic transitions was used to determine a multipole mixing ratio $\delta$ for each transition, when possible. An example of a mixing ratio calculation is displayed in Fig. \ref{fig:DeltaFit}. More than one mixing ratio may be reported for a given inelastic transition. This is because the curve representing all possible values of $\delta$ loops over itself in the asymmetry phase space, often yielding two solutions for each intersection. Branching ratios were also calculated using intensity and mixing ratio information. The lines associated with inelastic branches may be clearly visible at some detector positions, but may interfere with escape peaks in the spectra of others. Therefore, both mixing and branching ratios are reported only when permitted by such considerations. In particular, branching ratios are not reported above 6.4 MeV in the present dataset due to large level densities. Each calculation involving angular correlations used the formalism of Ref. \cite{Iliadis2021}. The reader is referred to Ref. \cite{Johnson2025} for a list of the experimental asymmetry quantities used to determine $J^\pi$ values. Spectra visually demonstrating the asymmetry of two elastic transitions at 5.05~MeV beam energy are presented in Fig.~\ref{fig:spectra_sample}; at $\theta=90^\circ$, one $M1$ transition is visible in detectors parallel to the beam polarization (Fig.~\ref{fig:spectra_sample} a) and an $E1$ one can be seen in the perpendicular position (Fig.~\ref{fig:spectra_sample} b), while both transitions are resolved by detectors at backward angles (Fig.~\ref{fig:spectra_sample} c and d).

A LaBr$_3$ scintillator, placed at a forward angle and facing a Cu scatterer downstream of the target, was used to monitor the photon beam flux. The energy profile of the beam was monitored with a HPGe detector moved directly into the highly-attenuated beamline at the start and end of each beam energy setting. Together, these two beam monitoring detectors give the energy-dependent photon flux $\Phi_\gamma(E)$ for the duration of each measurement. The latter was used to calculate the energy-integrated ground-state scattering cross section for each excited state,
\begin{equation}
    I_{s,0} = g(\frac{\pi\hbar c}{E})^2\frac{\Gamma_0^2}{\Gamma},
\end{equation}
where the spin factor $g=(2J+1)/(2J_0+1)$ depends on the excited-state spin $J$ and ground-state spin $J_0$, $E$ is the excited-state energy, and $\Gamma$ and $\Gamma_0$ are the level width and ground-state partial width, respectively \cite{Zilges2022}. Experimentally, this quantity is obtained from the intensity of the ground-state scattering peak in each detector $\mathcal{I}_{s,0}=A\cdot\Phi_\gamma(E)I_{s,0} W(\theta,\phi)$. The scaling factor $A$, which incorporates target specifics and normalization, was fit using data from five excited states with $\Gamma_0^2/\Gamma$ or $(2J+1)\Gamma_0^2/\Gamma$ values reported in Ref. \cite{Metzger1972}, which were also observed in the present dataset. These are the levels at 3346, 3617, 4338, 4462, and 4500~keV. Note that for the 3617-keV transition, the necessary $J^\pi$ quantum numbers were unambiguously determined in the present work, enabling the usage of this data point. For these cross-section calculations, excited states were assumed to have spin $J=1$. In some cases, particularly in the high-level-density region at the highest beam energies, the parity remains ambiguous and cross section values were reported for both the $E1$ and $M1$ decay scenarios.

\subsection{$\gamma-\gamma$ Coincidence Measurements}

The coincidence measurements, aimed at unveiling details of the level structure not apparent in the NRF scan, were performed at beam energies just below the  $^{68}$Zn proton-emission threshold of 9.98~MeV. The highest level density is expected in this energy regime, increasing the probability of observing inelastic transitions. Two different beam energies were used to avoid the possibility of devoting all counting time to a region where unforeseen nuclear structure effects or a low cross section might occur. Data was taken for 34 h at 9.46~MeV and 30 h at 9.79~MeV.

Data was written to file only if two or more detectors fired simultaneously within a larger time window encompassing several beam pulses. Post-processing defined true coincident signals as occurring within 100 ns of one another, whereas the beam pulse period is 179 ns. A sample of time-random coincidence events was generated using events outside this time window, which were then subtracted from the true coincidence events sample. A detailed description of the data acquisition and processing can be found in Ref. \cite{Johnson2025}.

Two-dimensional coincidence matrices were generated for all pairings of detector types; i.e. symmetrized matrices for clover-clover and CeBr$_3$-CeBr$_3$ events, and unsymmetrized ones for clover-CeBr$_3$ events. The energy calibration utilized room background lines and a strong ground-state transition from the 3346-keV level. Above the 3346-keV line, there were very few distinct or identifiable peaks in the total projection spectra used for calibration.

The $\gamma-\gamma$ coincidence matrices were analyzed by gating on specific $\gamma$ rays and examining the spectra they returned. Coincidence gating was performed using the \textsc{Cubix} software package, which allows the placement of separate gate and background regions, automatically scaling and subtracting the latter from the former \cite{Dudouet2024}.

\section{Results}

\subsection{Properties of NRF states}

\begin{figure}
    \centering
    \includegraphics[width=\linewidth]{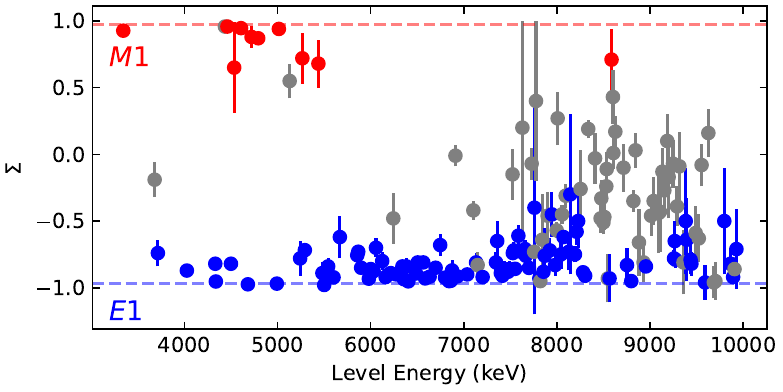}
    \caption{Asymmetry $\Sigma$ of $^{68}$Zn NRF ground state decays, measured by clover detectors at $\theta=90^\circ$. Levels assigned $J^\pi=1^+$ are plotted in red, while those assigned $J^\pi=1^-$ are shown in blue. Excited states from the NRF scan that lack firm $J^\pi$ assignments are plotted in gray. Note that the $J^\pi$ assignments take the backward-angle asymmetry into account, as well (not plotted here). Expected asymmetries of $M1$ and $E1$ transitions, with corrections for solid-angle effects, are shown as dashed lines in red and blue, respectively.}
    \label{fig:Asym}
\end{figure}

The NRF scan was carried out at 34 energies between 2.90 and 9.79~MeV, as described above, and identified 158 low-spin excited states within this energy range. These levels are listed in Table \ref{tab:levels_and_decays}, together with the measured transition energies and decay properties. Specifically, the proposed $J^\pi$ spin-parity assignments are based on the measured anisotropies reported in Ref. \cite{Johnson2025}. In Fig.~\ref{fig:Asym}, the marked difference between the anisotropies for $\gamma$ rays of $E1$ and $M1$ character  for all the transitions of Table \ref{tab:levels_and_decays} illustrates the power of the technique. These anisotropies were also instrumental in establishing the multipole mixing ratios $\delta$ and the branching ratios for transitions to excited states. As can be seen from Table \ref{tab:levels_and_decays}, most $J^\pi$ assignments are firm up to an excitation energy of $\sim7.5$~MeV. At higher excitations, the uncertainty in such assignments to individual levels increases due to the higher level density and the reduced resolving power of the array. 

\begin{longtable*}{ l c S S l c c}\\
\caption[$^{68}$Zn levels and decays from the NRF scan]{Level energies reached by NRF $E_{level}$, spin and parity $J^\pi$ of the state based on asymmetries given in Ref. \cite{Johnson2025}, transition energies $E_\gamma$ (elastic and inelastic decay), branching ratios and measured multipole mixing ratios $\delta$, as well as energies $E_f$ and spin-parity $J^\pi_f$ of the states reached in the decay.}\\\hline
\multicolumn{1}{c}{$E_{level}$ (keV)} & \multicolumn{1}{c}{$J^\pi$} & \multicolumn{1}{c}{$E_{\gamma}$} & \multicolumn{1}{c}{Branching Ratio ($\%$)} & \multicolumn{1}{c}{$\delta$$^\textit{a}$ } & \multicolumn{1}{c}{$E_f$} & \multicolumn{1}{c}{$J^\pi_f$}\\
\endfirsthead
\multicolumn{7}{c}
{\tablename\ \thetable\ -- \textit{Continued from previous page}} \\
\hline
\multicolumn{1}{c}{$E_{level}$ (keV)} & \multicolumn{1}{c}{$J^\pi$} & \multicolumn{1}{c}{$E_{\gamma}$} & \multicolumn{1}{c}{Branching Ratio ($\%$)} & \multicolumn{1}{c}{$\delta$$^\textit{a}$} & \multicolumn{1}{c}{$E_f$} & \multicolumn{1}{c}{$J^\pi_f$}\\\hline
\endhead
\hline \multicolumn{7}{r}{\textit{Continued on next page}} \\
\endfoot
\hline
\endlastfoot
\hline
\input{Results/levels_and_decays_edits}
\label{tab:levels_and_decays}
\end{longtable*}
\noindent$^\textit{a}$ All mixing ratios consistent with the data are listed. It should be noted, however, that mixing ratios with a smaller absolute value indicate dominance of the lower multipole and are more likely than larger-valued ones.\\
$^\textit{b}$ From Ref. \cite{McCutchan2012}.\\
$^\textit{c}$ Calculated using only detectors at backward angles (see Ref. \cite{Johnson2025} for details).

\begin{figure*}
    \centering
    \includegraphics[width=0.8\linewidth]{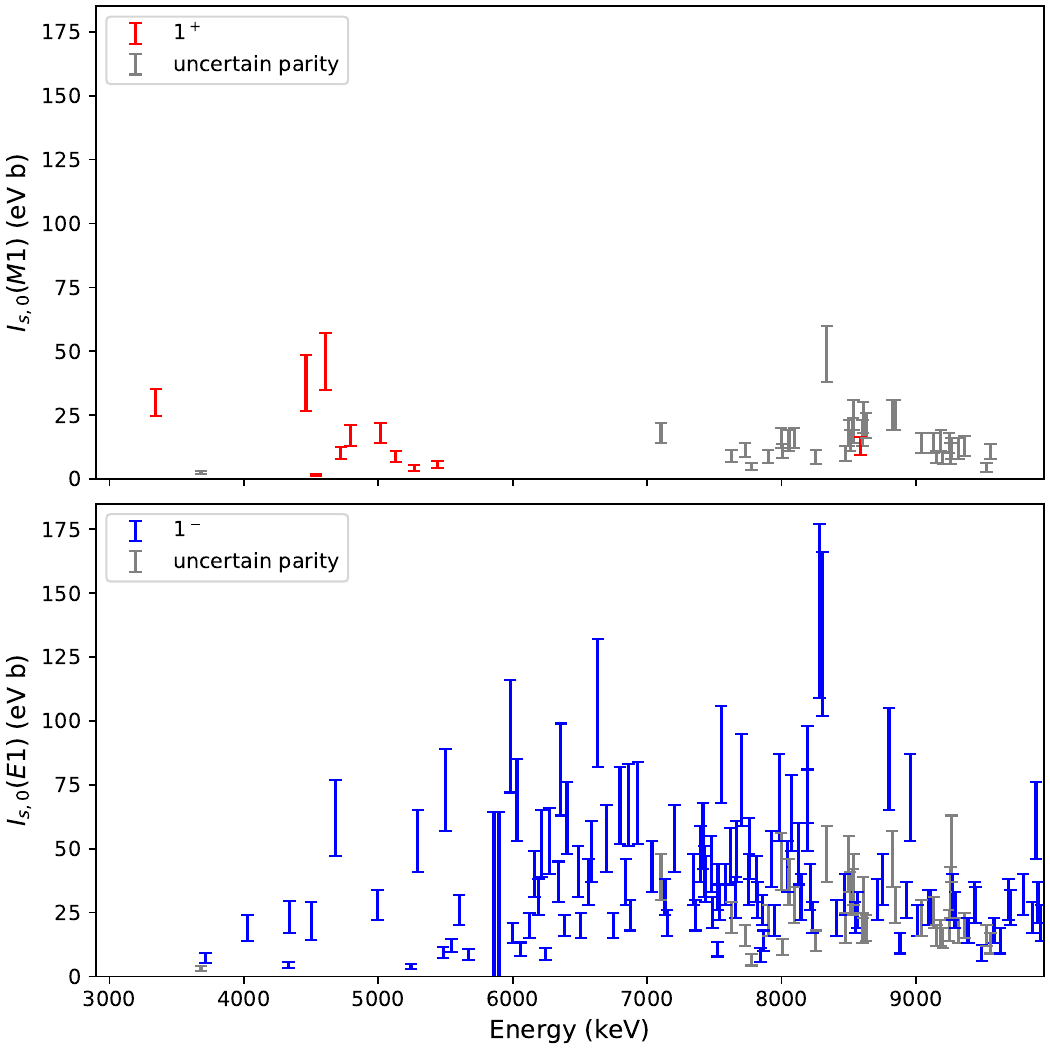}
    \caption{Elastic scattering cross section $I_{s,0}$ for all discrete NRF states studied in the scan. The top panel includes $M1$ transitions in red and the bottom panel $E1$ transitions in blue. Levels with unknown parity are plotted in gray in both panels.}
    \label{fig:dist_spin1}
\end{figure*}

Scattering cross sections are presented in Table \ref{tab:CrossSection} and displayed in Fig.~\ref{fig:dist_spin1}. The values are inferred from the data assuming  a $J=1$ spin for each state. In instances where the parity is uncertain, cross sections are quoted for both $1^+$ and $1^-$ possibilities. This uncertainty is also reflected in Fig.~\ref{fig:dist_spin1}, where contributions of levels with unknown parity are plotted in gray. Due to an erroneous trigger setting at the 5.85-MeV beam energy, flux data was not available. Therefore, cross sections could not be calculated for the 5856-, 5867-, and 5898-~keV states excited at this energy, and no cross sections are listed in Table \ref{tab:CrossSection}. However, for plotting purposes only, the $E1$ strength was estimated by averaging the total strength at neighboring beam energy settings. Note that no $M1$ transitions were identified at this 5.85-MeV beam energy.

\begin{longtable}{ c c c }\\
\caption{Ground-state scattering cross section values $I_{s,0}$ for all levels studied in the present NRF scan, assuming $J=1$ spin values. When the parity $\pi$ is uncertain, values are given for both $\pi=-$ and $\pi=+$, respectively.}\\\hline
\multicolumn{1}{c}{$E_{level}$ (keV)} & \multicolumn{1}{c}{$J^\pi$} & \multicolumn{1}{c}{$I_{s,0}$ (eV b)} \\\hline
\endfirsthead
    \hline
    \multicolumn{1}{c}{$E_{level}$ (keV)} & \multicolumn{1}{c}{$J^\pi$} & \multicolumn{1}{c}{$I_{s,0}$ (eV b)} \\\hline
    \endhead
    \endfoot
    \hline
    \endlastfoot
    \hline
\input{Results/CrossSectionsEditedEnergies}
\label{tab:CrossSection}
\end{longtable}\noindent$^\textit{a}$ Calculated using $\Gamma_0^2/\Gamma$ values from Ref. \cite{Metzger1972}.

\subsection{Coincidence Relationships}
\begin{figure*}
    \centering
    \includegraphics[width=\linewidth]{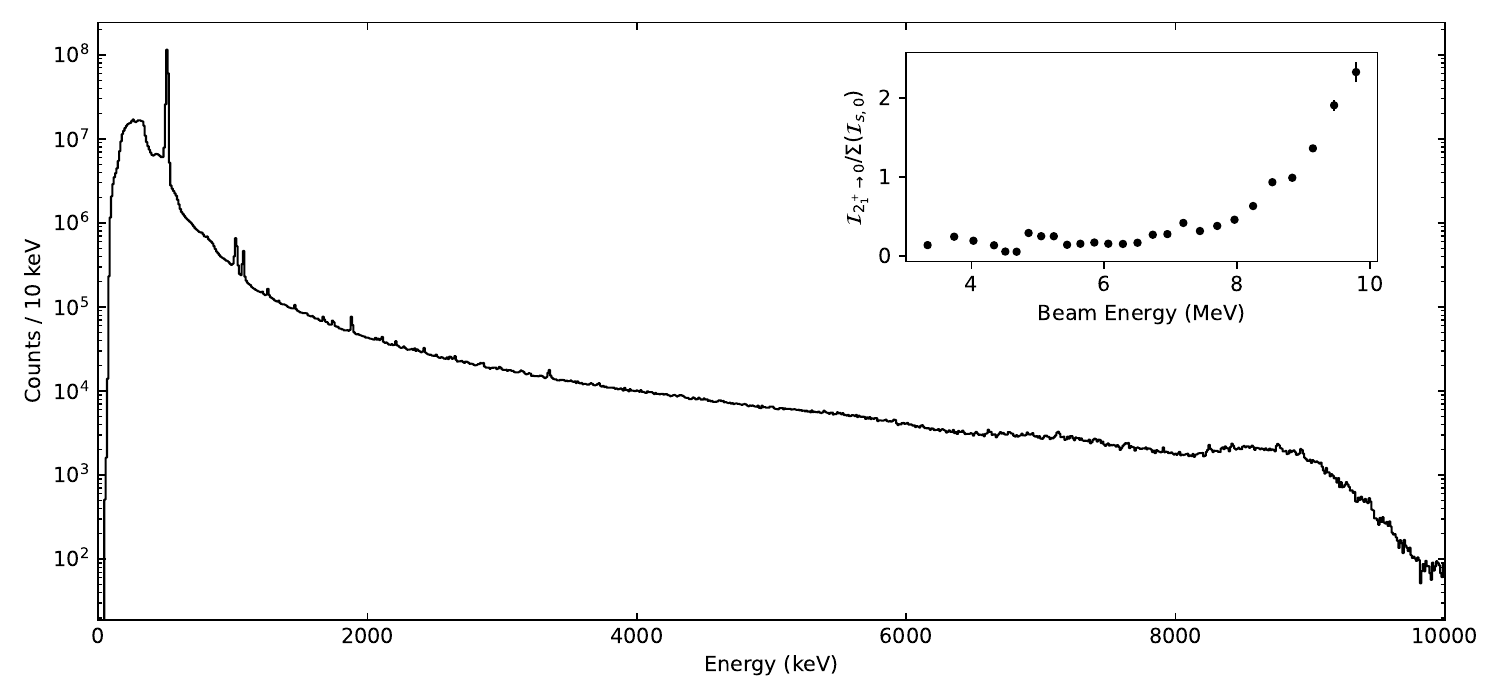}
    \caption{Total projection onto one energy axis of the clover-clover coincidence matrix at a beam energy of 9.46~MeV. The inset provides the ratio of the intensity of the $2^+_1\rightarrow0^+_1$ ground state transition to that of all spin-1 ground-state transitions by beam energy. Note that for this plot, no correction for angular distributions has been made.}
    \label{fig:TotalProjClover}
\end{figure*}

As stated above, coincidence events between all the detectors in the Clover Array were recorded at beam energies of 9.46 and 9.79~MeV. The inset in Fig.~\ref{fig:TotalProjClover} presents a ratio of the intensity of the $2^+_1\rightarrow 0^+_1$ transition versus that of all spin-1 ones as a function of beam energy. The intensity of the ground state transition is viewed as a proxy for the total inelastic intensity; i.e. this $2^+_1$ level is expected to act as a funnel for deexcitation from higher-lying states. It is apparent that inelastic decay paths play a dominant role as the excitation energy increases, particularly above 8~MeV, close to the particle-emission threshold. This observation justifies the choice of beam energies for the coincidence measurements. 

\begin{figure}
    \centering
    \includegraphics[width=\linewidth]{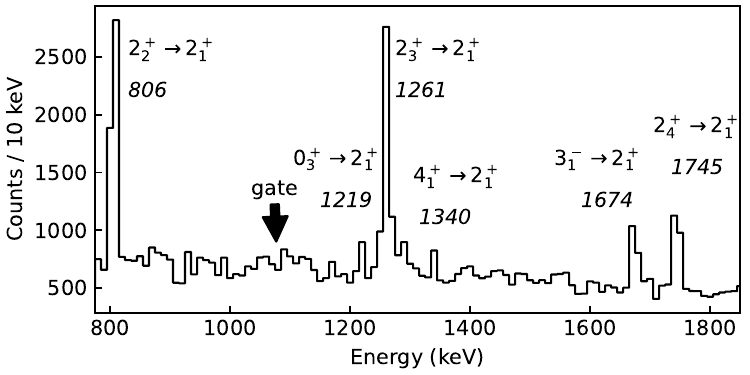}
    \caption{Clover spectrum after gating on the 1077-keV $\gamma$ ray corresponding to the ground-state decay from the $2^+_1$ level. Several coincident $\gamma$ rays have been identified.}
    \label{fig:CloverCoinc1077Gate}
\end{figure}

\begin{figure}
    \centering
    \includegraphics[width=\linewidth]{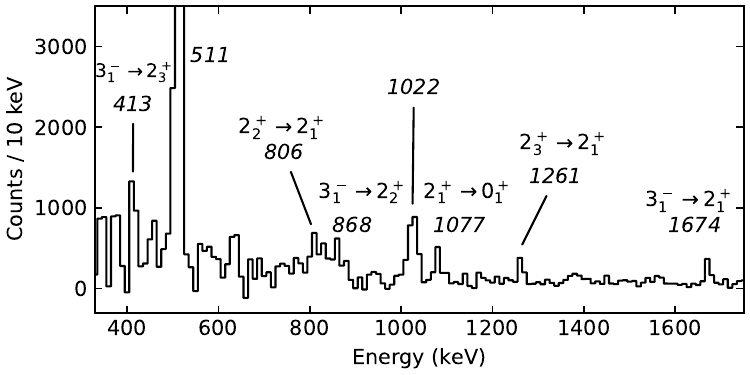}
    \caption{Clover-clover coincidence spectrum for a gate on the transition from the $4^+_3$ state to the $3^-_1$ level. Three decay branches from the $3^-_1$ level are visible in the gated spectrum.}
    \label{fig:CloverCoinc835Gate}
\end{figure}

A total projection of the coincidence matrix recorded at 9.46~MeV is found in Fig.~\ref{fig:TotalProjClover}. Figure \ref{fig:CloverCoinc1077Gate} presents the spectrum in coincidence with the  1077-keV, $2^+_1\rightarrow 0^+_1$ transition where a number of $\gamma$ rays associated with the feeding of the lowest excited state is visible, illustrating the sensitivity achieved in the experiment. Coincidence relationships for a weaker decay branch are illustrated in Fig.~\ref{fig:CloverCoinc835Gate}, where a gate is placed on the 836-keV, $4^+_3\rightarrow3^-_1$ transition. Several decay paths out of the $3^-_1$ level toward all lower-lying $2^+$ states are visible, including a branch to the $2^+_2$ state which was not reported in prior work \cite{McCutchan2012}. Furthermore, the secondary deexcitation $\gamma$ rays of each $2^+$ level are also captured in this gate. 

\begin{figure}
    \centering
    \includegraphics[width=\linewidth]{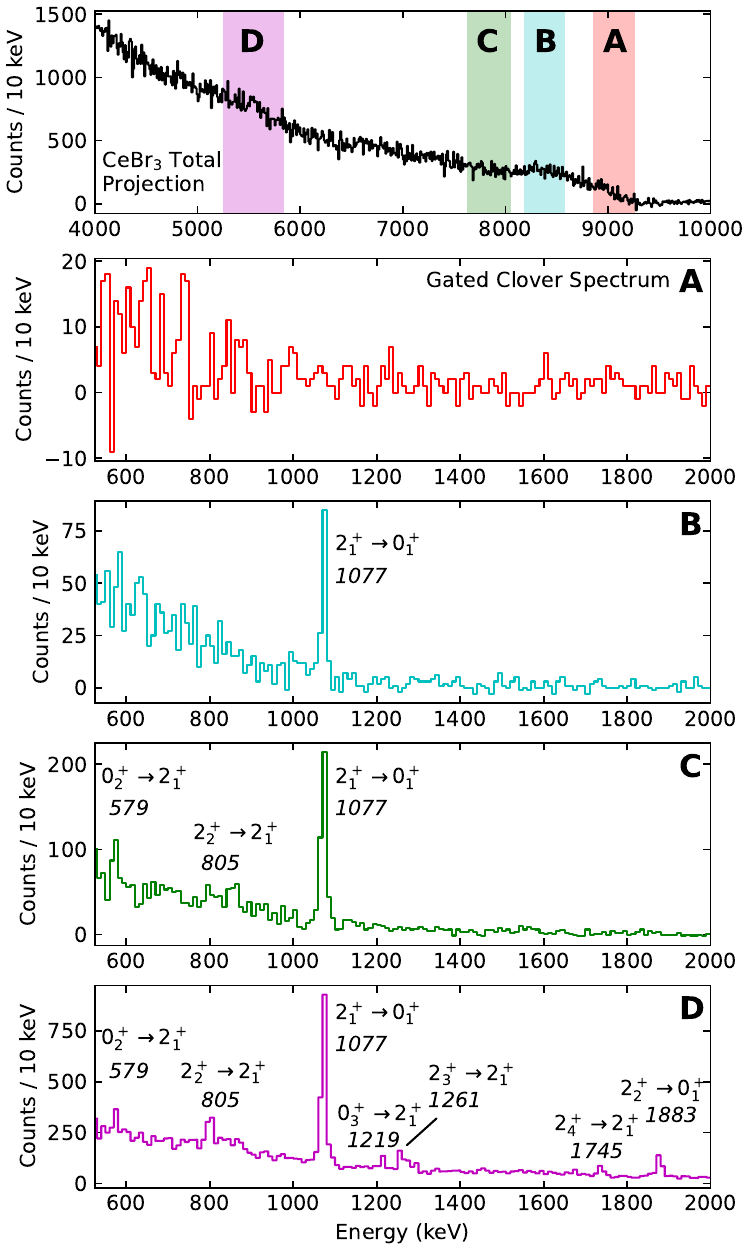}
    \caption{Clover spectra in coincidence with various high-energy gates on the CeBr$_3$ detector response (top panel) for the 9.46-MeV beam energy. Signatures of levels populated directly by the high-energy spin-1 states are visible in the gated clover spectra A-D in the second through fifth panels. See text for details. }
    \label{fig:CeBr3_bumps}
\end{figure}

As shown in Ref. \cite{Johnson2025}, the analysis of coincidence relationships between CeBr$_3$ scintillators and HPGe clover detectors confirmed the placement of the various transitions. This is illustrated in Fig.~\ref{fig:CeBr3_bumps} for events obtained at a 9.46-MeV beam energy.  The coincidence gates placed on broad energy ranges in CeBr$_3$ spectra exhibit strong correlations with the low-energy transitions observed in the HPGe clover detectors. Gate A in the CeBr$_3$ total projection (Fig.~\ref{fig:CeBr3_bumps}, top panel) is placed at an energy only a few hundred~keV below the nominal beam energy. With the $2^+_1$ state being located at 1077~keV, the corresponding gated HPGe spectrum (Fig.~\ref{fig:CeBr3_bumps}, second panel) is essentially empty, containing only scattering events and no inelastic transitions. Gate B, which covers a 0.2~MeV-wide energy range centered around 8.4~MeV (i.e., $E_{beam} - 1077$~keV), demonstrates that the transitions involved are associated with the direct feeding of only the $2^+_1$ level (see third panel Fig.~\ref{fig:CeBr3_bumps}). By moving the CeBr$_3$ gate further down in energy to 0.2~MeV- and 0.3~MeV-wide ranges centered around 7.8 and 5.5~MeV, gates C and D in the 4th and 5th panels, transitions associated with decays from higher-lying yrast and near-yrast states become apparent, herewith validating their placement in the level scheme.

\begin{figure}[]
    \centering
    \includegraphics{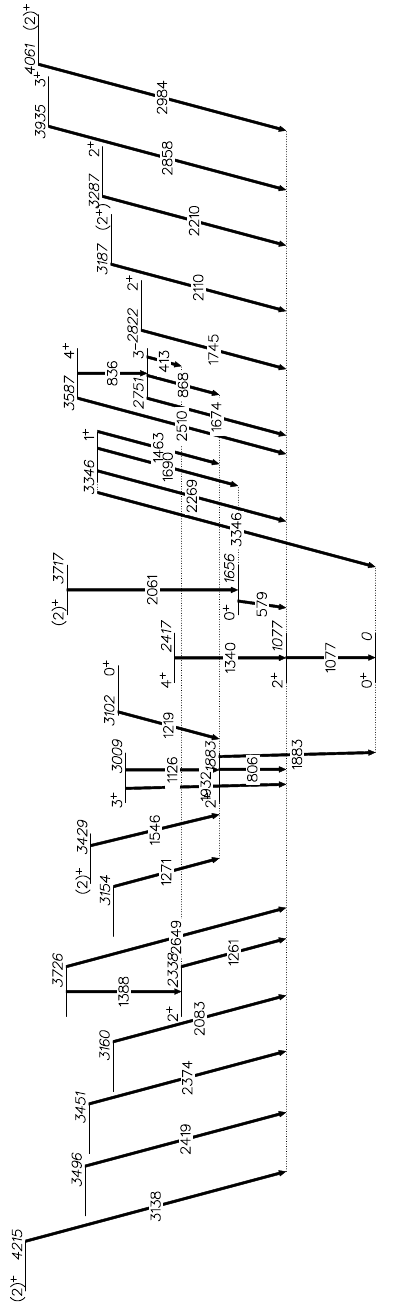}
    \caption{Level scheme obtained from the $(\gamma,\gamma')$ coincidence analysis.}
    \label{fig:levelScheme}
\end{figure}

The final level scheme deduced from this experiment can be found in Fig.~\ref{fig:levelScheme} where the NRF technique is shown to have populated discrete states up to an excitation energy of 4.2~MeV while reaching levels with spins as high as $J = 4$. The scheme contains 23 excited states linked by a total of 32 transitions.  All the levels have been reported in earlier studies \cite{McCutchan2012} although, in some instances, the present data have allowed a more precise determination of the excitation energy and more restrictions on possible spin-parity assignments. Two of the observed linking transitions are new: the 1690-keV $\gamma$ ray linking the 3346-keV $1^+_1$ and 1656-keV $0^+_2$ states, and the 868-keV one associated with the $3^-_1\rightarrow2^+_2$ decay mentioned above.  

\section{Discussion}

The results presented above considerably expand the existing nuclear structure data for $^{68}$Zn, especially regarding low-spin states. As indicated above, the NRF scan was able to identify and characterize the properties of 158 excited states, thereby greatly adding to the earlier NRF studies of Refs. \cite{Moreh1983} and \cite{Metzger1972}. Firm $J^\pi$ spin-parity assignments are proposed for many of the states reported in Table \ref{tab:levels_and_decays}. For the others, some possible $J^\pi$ values found in the literature \cite{McCutchan2012} could be ruled out based on the present data. For many of the observed levels, deexcitation paths to several low-lying excited states were identified as well, and multipole mixing ratios and branching ratios were determined. Scattering cross section values were also quantified with an experimental sensitivity spanning two orders of magnitude: the smallest scattering cross section of 1.6(4)~eV~b is reported for the $1^+$ state at 4535~keV, and the largest, 143(34)~eV~b for the $1^-$, 8282-keV level. In this context, it should be noted that a level at 7359~keV\footnote{Ref. \cite{Moreh1983} lists this level at 7362~keV.}  had been reported in the earlier work of Moreh \textit{et al.} \cite{Moreh1983} as decaying through one of the strongest $E1$ transitions observed in this entire region of the nuclear chart. However, the present cross section for this state was determined to be 24(6)~eV~b (see Table \ref{tab:CrossSection}), more than an order of magnitude smaller than the previous value, herewith ruling out a special importance for this excitation. 

As is clear from Fig.~\ref{fig:dist_spin1}, the $E1$ strength dominates the $M1$ one over the entire range of excited levels, as is common for nuclei in this mass range.  These data provide an opportunity to perform extensive comparisons with calculations carried out with some of the up-to-date effective interactions developed for nuclei in the $A = 50-80$ region of the nuclear chart.

\subsection{$E1$ and $M1$ Strengths: Shell-Model Interpretation}

Shell-model calculations were performed with the \textsc{NuShellX@MSU} code \cite{Brown2014}. With $^{68}$Zn being two protons beyond the $Z=28$ shell closure and two neutrons below the $N=40$ subshell one, initial calculations were carried out within the $jj44$ model space. The latter consists of an inert $^{56}$Ni core and the $0f_{5/2},1p_{3/2}, 1p_{1/2},$ and $0g_{9/2}$ single-particle states. Two effective inter-nucleon interactions, JUN45 and $jj44b$, were used in this model space. The former interaction is fit to data for 69 nuclei between $A = 63 - 96$, but excludes Ni and Cu isotopes in an attempt to understand the impact of the missing $0f_{7/2}$ orbital and the softness of the $^{56}$Ni core. Nuclear data for several Zn isotopes, including $^{68}$Zn, were included in the fit, which is expected to perform best for nuclei with $Z\approx32$ and $N\approx50$ \cite{Honma2009}. In contrast, the parameters of the $jj44b$ effective interaction were fit to data for the Ni and Cu isotopic chains, as well as data in the region near $N = 40$ \cite{Mukhopadhyay2017}.

\begin{figure*}
    \centering
    \includegraphics[width=0.7\linewidth]{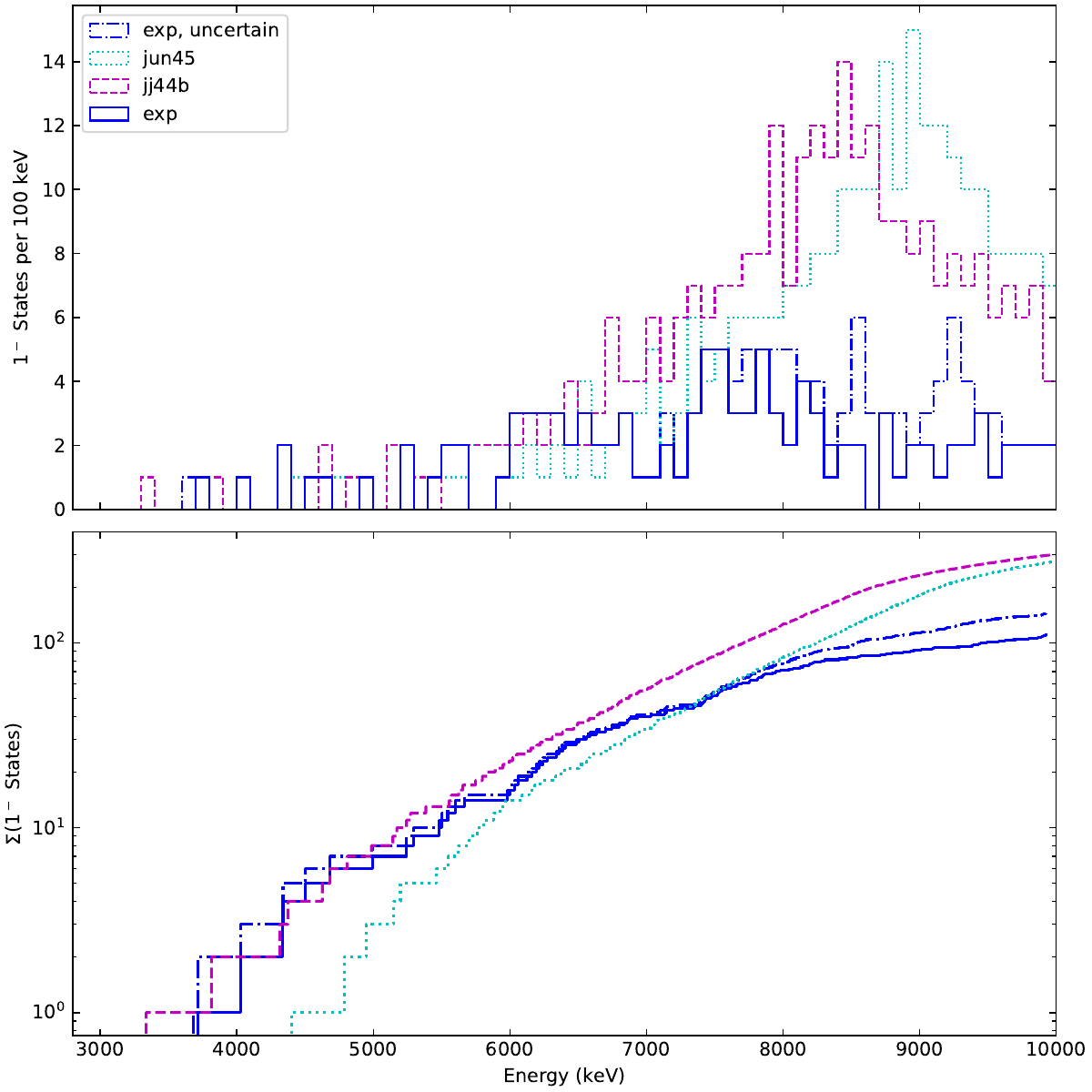}
    \caption{Distribution of $1^-$ states by energy as measured experimentally (solid dark blue, and including states with unknown parity in dash-dotted dark blue), and in the shell model with the $jj44$ model space ($jj44b$ effective interaction in dashed purple and JUN45 in dotted light blue). The top panel shows the number of states per 100~keV, and the bottom panel displays the running sum of the number of $1^-$ states.}
    \label{fig:e1_density}
\end{figure*}

As shown below, the shell-model calculations are capable of predicting energies and strengths of the $1^+$ states. For the $1^-$ levels, however, particle-hole states giving rise to $E1$ transitions lie outside the $jj44$ model space. Yet, this particle-hole $E1$ strength mixes with that of the many-particle $1^-$ states within the $jj44$ model space. At present, there is no theoretical method to calculate this mixing and, as a result, scattering cross sections, branching, and mixing ratios have not been calculated.  Only the distribution of states is presented in the two panels of Fig.~\ref{fig:e1_density} for comparison between experiment and theory. The top of Fig.~\ref{fig:e1_density} displays the distribution of the $1^-$ states as a function of energy. Good agreement is found between the experiment and shell-model calculations below $\sim6.5$~MeV. At higher excitation energy the experimental levels are found to be fewer in number than predicted. For example, around 8~MeV, the shell-model calculations predict level densities of up to fifteen $1^-$ states per 100~keV, while only up to six levels are observed. In this high-energy domain, the calculated $1^-$ states can be too close in excitation energy to be resolved. As a result, their identification as well as their strength may be below the detection limit of this experiment. Nevertheless, the trends in the data and the calculations are in general agreement. This is perhaps best visualized in the bottom panel of Fig.~\ref{fig:e1_density} which compares a running sum of experimental and calculated $1^-$ states as a function of excitation energy. Calculations with both interactions reproduce the general trend seen in the data, although the $jj44b$ interaction predicts a level density closer to experiment below 5.5~MeV. The $1^-$ levels in JUN45 are computed to start at an excitation energy about one~MeV higher when compared to $jj44b$ calculations. In addition, the latter compute a less steep rise with excitation energy than that predicted by the former and mirrors the data better.

\begin{figure*}
    \centering
    \includegraphics[width=0.8\linewidth]{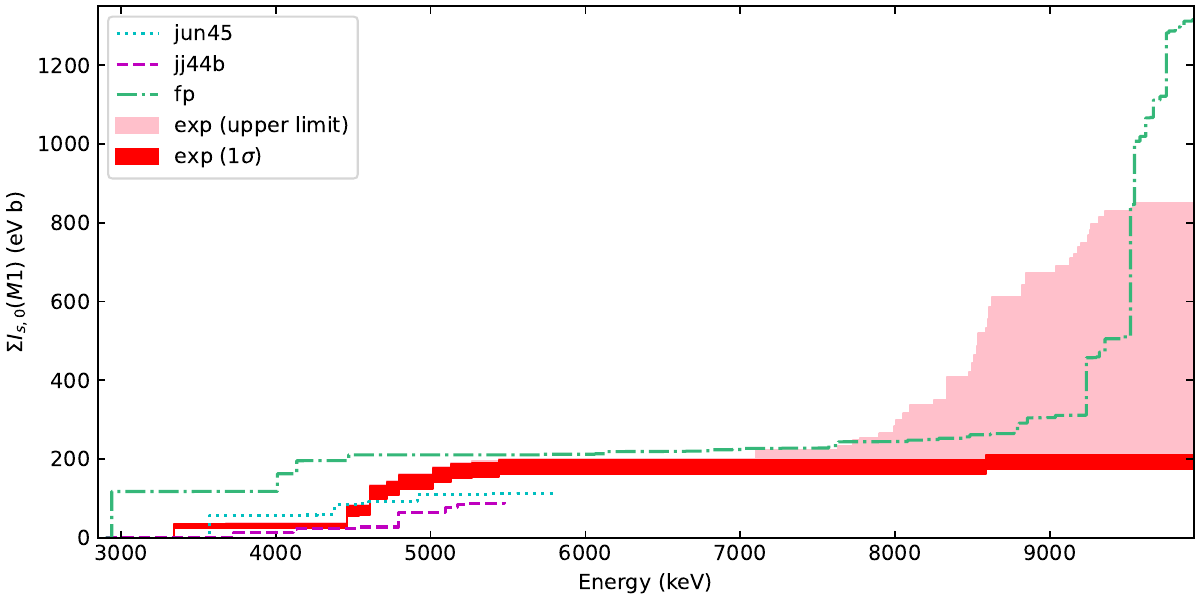}
    \caption{Experimental (red) running sum of the scattering cross section for $1^+$ states compared to shell-model calculations of the same quantity in the $jj44$ model space with the JUN45 (light blue, dotted line) and $jj44b$ (purple, dashed line) effective interactions. Also included are calculations in the $fp$ model space (green, dot-dashed line) which use the GXPF1A effective interaction.}
    \label{fig:Is_M1}
\end{figure*}

For the $1^+$ levels, the reservations expressed above about comparisons between calculated and measured cross sections do not apply. In the $jj44$ model space, the $M1$ strength for these levels comes from $1p_{1/2} - 1p_{1/2}$, $1p_{3/2} - 1p_{3/2}$, $1p_{1/2} - 1p_{3/2}$, $0f_{5/2} - 0f_{5/2}$ and $0g_{9/2} - 0g_{9/2}$ single-particle contributions. The $jj44$ model space does not include the $0f_{7/2}$ and $0g_{7/2}$ orbitals and is therefore missing contributions coming from these orbitals. Importantly, excitation from the $0f_{7/2}$ shell to the $0f_{5/2}$ one gives rise to the $M1$ giant resonance which is missing in the $jj44$ model space. This shortcoming is discussed below with regard to the $fp$ model space. However, to some approximation, the mixing of the $M1$ resonance into other $M1$ transitions in the $jj44$ model space is accounted for by the use of an effective $M1$ operator. As was the case in Ref. \cite{Johnson2023}, the effective $M1$ operator from Ref. \cite{Honma2009} was used with the spin $g$ factor quenched by a factor of 0.7. The running sum of the $M1$ cross section $I_{s,0}$ is provided in Fig.~\ref{fig:Is_M1}, where it is compared with the results of each set of shell-model calculations. 

The general trend in the calculated $M1$ strength is reproduced satisfactorily by both the $jj44b$ and JUN45 Hamiltonians. Both interactions predict the presence of very few states below 4~MeV. At most two states are observed experimentally, with the only spin-1 level with definite parity being located somewhat lower in excitation energy than computed by both interactions\footnote{The $jj44b$ calculations predict a first $1^+$ state lower in energy than the experimental one. The associated cross section is, however, $3.2\times10^{-7}$~eV~b, well below the detection limit.}, and with a measured cross section intermediate between the two calculated ones (Fig.~\ref{fig:Is_M1}). By examining the calculated single-particle matrix elements for the dominant decay multipolarity of these states, the main orbitals involved can be identified. In the JUN45 calculations, the $M1$ strength arises from excitations within the neutron $0f_{5/2}$ shell, whereas the neutron $0g_{9/2}$ transitions are dominant in the $jj44b$ computations. Between 4 and 5~MeV excitation energies, most measured $M1$ strength is concentrated in a few fairly strong states, a feature found in the calculations as well. In this energy range, the calculated and measured running sums of cross sections exhibit a rise, with the $M1$ excitations calculated by JUN45 occurring at somewhat lower excitation energy than those computed within $jj44b$. Both calculations associate this strength with excitations involving mixed configurations within specific orbitals; i.e., the proton $1p_{3/2}$ and both the proton and neutron $0f_{5/2}$ shells for JUN45 and, for $jj44b$, the proton $0f_{5/2}$,$ 1p_{3/2}$ and $0g_{9/2}$ shells as well as the neutron $0f_{5/2}$ and $0g_{9/2}$ ones. Smaller contributions from transitions from the proton and neutron $1p_{1/2}$ to the $1p_{3/2}$ shells are also present for both calculations. Moving into the $\sim4.7 - 5.2$~MeV range, the experimental running sum shows that additional strength is steadily added. This is echoed in the calculations, as they also continue to rise in this region. Here, the $jj44b$ sum amounts to roughly 45\% of the experimental one, while that of JUN45 reaches 65\%. Above $\sim5.2$~MeV, the experimental running sum of scattering cross section data for the $1^+$ states appears to saturate at a roughly constant value which persists at least up to 7~MeV. This saturation is also present in the calculations and reflects the fact that the valence model space available for the particle-hole excitations responsible for these $1^+$ states has been exhausted. Hence, in Fig.~\ref{fig:Is_M1} the results of these calculations do not extend beyond 6~MeV.

\begin{figure*}
    \centering
    \includegraphics[width=0.8\linewidth]{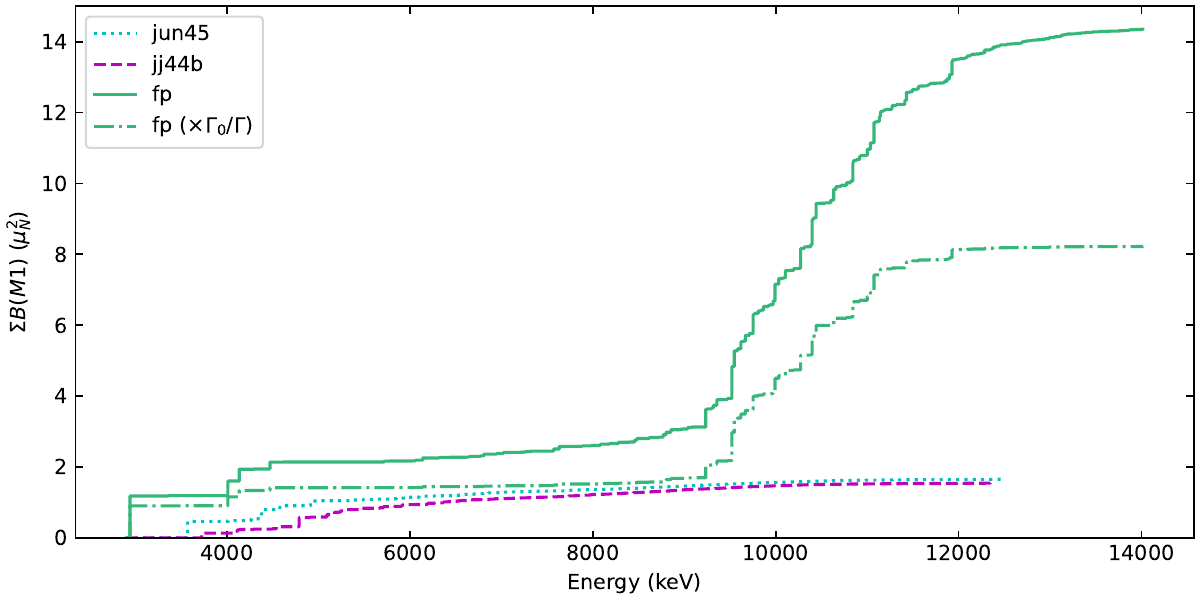}
    \caption{Running sum of the $B(M1)$ strength from shell-model calculations in both the $jj44$ model space with the JUN45 (blue) and $jj44b$ (purple) effective interactions as well as the $fp$ model space (green) with the GXPF1A effective interaction. The $B(M1)$ values modulated by the ground-state branching ratio $\Gamma_0/\Gamma$ are also shown (green, dot-dashed).}
    \label{fig:BM1}
\end{figure*}

To account for the additional $M1$ strength seen in the data above 6~MeV, excitations beyond the $jj44$ valence space had to be considered. Specifically, the impact of allowing excitations from the $0f_{7/2}$ states were investigated by considering the full $fp$ model space which consists of an inert $^{40}$Ca core and enables excitations in the $0f_{7/2}$, $1p_{3/2}$, $0f_{5/2}$ and $1p_{1/2}$ proton and neutron orbitals, but excludes the $\nu0g_{9/2}$  and $\pi0g_{9/2}$ single-particle states. This set of calculations in the full $fp$ model space used the GXPF1A effective interaction \cite{Honma2005} where parameters were fit to data for many nuclei with $Z = 20 - 40$ and $N = 32 - 50$, including the lighter-mass even-even Zn isotopes $^{60-64}$Zn, but not $^{68}$Zn. The running sum of the scattering cross sections for this calculation is given in green in Fig.~\ref{fig:Is_M1}. While calculations with this interaction do not reproduce the $M1$ cross sections at the lower energies, where the model space is inadequate, they capture the flattening of the running sum between 5.5 and 7.5~MeV, although the accumulated strength is overpredicted by $\sim20\%$. This overprediction is explained by a general overestimate of the occupation of the $p$ orbitals. Within the $fp$ model space, the $0g_{9/2}$ shell is inaccessible, leading to an overestimation of the contribution of spin-flip transitions between the $1p_{3/2}$ and $1p_{1/2}$ shells. Specifically, these $fp$ calculations result in strong population, below 4~MeV, of $1^+$ states mainly built of excitations from the $1p_{3/2}$ to $1p_{1/2}$ orbitals for protons and neutrons, respectively. Within the $jj44$ space, this $p$-shell occupation is reduced resulting in better agreement with the data. At higher energies, in the $5 – 7$~MeV range, the $1^+$ levels are more mixed, have smaller strengths and result in the small rise in cross section values exhibited by the data. The very large increase of strength computed to occur beyond 9~MeV is associated mainly with proton spin-flip excitations from the $0f_{7/2}$ shell to the $0f_{5/2}$ one. This spin-flip giant resonance is even more readily apparent in Fig.~\ref{fig:BM1}, where calculations of the $B(M1)$ strength are plotted beyond the particle emission threshold. 

The results from the present work can be compared with those reported recently by Ref. \cite{Kelly2026}, which combined one-neutron transfer, $(\gamma,\gamma')$, $(e,e’)$, and $(p,p’)$ reactions to infer that the overall $M1$ strength in semi-magic $N = 28$ $^{50}$Ti is
not dominated solely by neutron $0f_{7/2} - 0f_{5/2}$ transitions near the particle-emission threshold.  Calculations with the GPFX1A Hamiltonian used above find a peak in the $^{49}$Ti$(p,d)$ $0f_{5/2}$ transfer to $1^+$ states accompanied by a peak in the $B(M1)$ strength around 11.2~MeV, a value about 1.5~MeV higher than that found in the Ref. \cite{Kelly2026} experiment. The associated computed width of $\sim1$~MeV is smaller than that observed in the measurements. It is indicative of a spreading width not contained in the calculations and is coming from mixing with configurations beyond the $fp$ model space, as suggested in Ref \cite{Kelly2026}. In the present case of $^{68}$Zn, the calculations attribute the  $M1$ strength to  $0f_{7/2}$ to $0f_{5/2}$ proton excitations, and additional spreading may be anticipated as well.

While the $fp$ model space offers some advantages at higher excitation energies, it should be noted that it does not account for $E1$ excitations due to the lack of an intruder orbital with which to build them. None of the shell-model calculations can, by themselves, reproduce the observed NRF results for $^{68}$Zn. Thus, the latter give valuable input for new shell-model approaches in this region, either through the creation of new effective interactions, or through development methods utilizing expanded model spaces.

\subsection{Shell-Model Interpretation of Levels Near the $^{68}$Zn Ground State}

The shell-model calculations described above can be tested further by carrying out comparisons with the properties of levels close to the $^{68}$Zn ground state.  As shown in Fig.~\ref{fig:levelScheme}, the analysis of the coincidence events indicated that a total of 23 excited states was populated from levels close to the particle emission threshold, and that many transitions linking these were observed. Compared to the latest compilation of Ref. \cite{McCutchan2012}, two new decay paths ($3^-_1\rightarrow2^+_2$ and $1^+_1\rightarrow0^+_2$) were established. Furthermore, known levels up to $J^\pi = 4^+$ were confirmed, underscoring the sensitivity of the present technique. 

\begin{figure*}
    \centering
    \includegraphics[width=\linewidth]{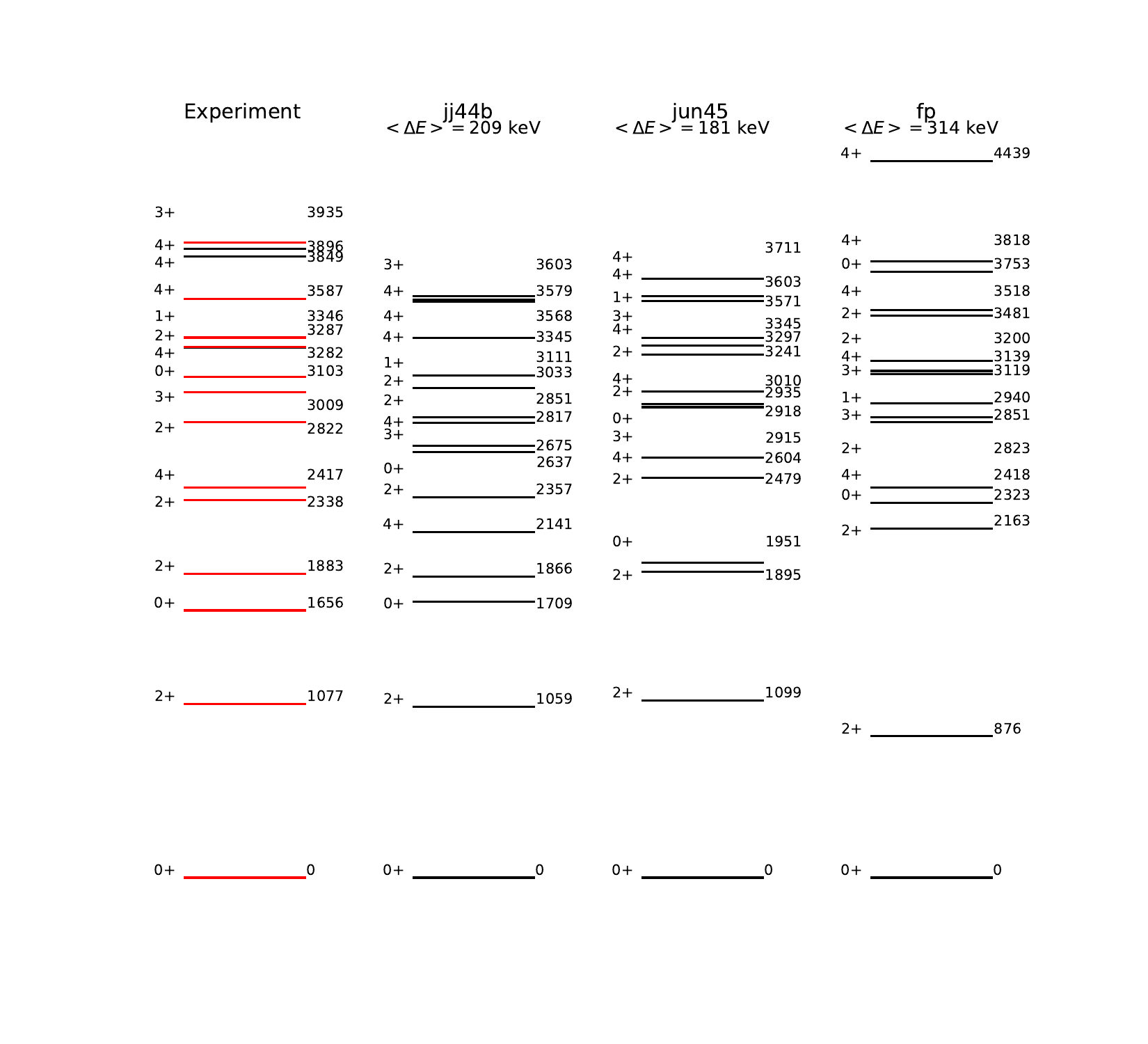}
    \caption{Distribution of positive-parity states in $^{68}$Zn with spin up to $J^\pi=4^+$, along with their shell-model counterparts. Only levels with firm $J^\pi$ assignments are considered. Levels seen in the present measurements are highlighted in red, with only a few additional states taken from Ref. \cite{McCutchan2012}. The average energy deviation is given for each set of shell-model calculations.}
    \label{fig:SM_lowE}
\end{figure*}

Figure \ref{fig:SM_lowE} displays all the known positive-parity $^{68}$Zn states with firm spin assignments below $J=4$ and up to 4~MeV, together with their shell-model counterparts calculated within the two model spaces with the three Hamiltonians described above.  The average energy deviation for each set of calculated levels is also given in the figure. The smallest deviation is achieved in the calculations using the $jj44$ model space, with 209~keV for the $jj44b$ calculations and 181~keV for the JUN45 ones. Both values are in line with expectations for shell-model calculations reproducing data in this mass region. The agreement between experiment and calculations within the $jj44$ space is also readily apparent from the ability to reproduce general trends such as the overall sequence of $J^\pi$ values as a function of excitation energy, or the clustering of the $0^+_2$ and $2^+_2$ states and that of the $2^+_3$ and $4^+_1$ levels. Calculations in the $fp$ model space predict the excited states to span a broader energy range than those seen experimentally, hence the larger energy deviation of 314~keV. As discussed earlier, the $fp$ calculations appear to be inadequate for describing the level structure close to the ground state, herewith providing an indication of the importance of the inclusion of the $0g_{9/2}$ orbital in the model space. A further comparison, focused solely on the $2^+$ levels below 4~MeV (see Fig. 5.5 in Ref. \cite{Johnson2025}), reveals a remarkable correspondence between experimental levels and the $jj44b$ calculations, with an average energy deviation of only 37~keV. The JUN45 interaction also performs well, with a deviation of only 176~keV originating from a slight systematic over-prediction of the energy of the $2^+$ excitations above 2.4~MeV. The two interactions in the $jj44$ model space predict slight differences in orbitals involved in the $2^+$ excitations in this region, but do not point to a dominant configuration which would drive the energy difference. The $fp$ calculations overpredict the energy of the $2^+$ excitations more severely (by 600~keV), likely because of the inadequacy of the model space as discussed earlier.

Some excited states were observed in the coincidence data to deexcite through several branches to other low-lying levels, offering the possibility to test the calculations further by comparisons between the branching ratios. This includes both the $1^+_1$ and $3^+_1$ levels, and the newly-observed decays of the $0^+_3$ and $2^+_2$ states. The relevant experimental branching ratios as well as those predicted by the shell-model calculations are presented in Table \ref{tab:branchings}.  It is clear that the $fp$ calculations do not predict the ratios accurately for any of the cases under consideration. Transitions between these levels in the latter calculations often involve the $0f_{7/2}$ orbitals for both protons and neutrons which involve the breaking of the $^{56}$Ni core, and have been shown above to be involved only at higher excitation energies. For the four selected excited states of Table \ref{tab:branchings}, the JUN45 effective interaction in the $jj44$ model space predicts the branching ratios the best. The $2^+_2$ level deexcites to both the ground state and the $2^+_1$ level, with the strongest branch being the former. Each calculation predicts the opposite, placing the decay to the $2^+_1$ state as the dominant branch. Nevertheless, the JUN45 effective interaction gives the branching ratios closest to those found experimentally and identifies the ground state decay with proton transitions within the $1p_{3/2}$ shell and between the $1p_{3/2}$ and $1p_{1/2}$ ones, as well as neutron transitions between the $0f_{5/2}$ and $1p_{1/2}$ shells. The branch to the $2^+_1$ level is more mixed and involves the same transitions, with the addition of those between the proton $0f_{5/2}$ and $1p_{3/2}$ orbitals and neutron excitations within the $0f_{5/2}$ and $0g_{9/2}$ shells as well as between the $1p_{3/2}$ and $1p_{1/2}$ ones. 

\begin{table}[]
    \centering
    {\renewcommand{\arraystretch}{1.5}
    \caption{Branching ratios of selected $^{68}$Zn excited states. Experimental values  (``exp.") for the $2^+_2$, $3^+_1$, and $0^+_3$ levels are from Ref. \cite{McCutchan2012}, while those of the $1^+_1$ state are from the present work, with the addition of the intensity of the $2^+_1$ branch from Ref. \cite{McCutchan2012}, which was not seen in the NRF scan but was observed in the coincidence measurements.}
    \begin{tabular}{c c c S S S}
    \hline\hline $J^\pi_i$&  $J^\pi_f$&\multicolumn{4}{c}{Branching Ratio (\%)}\\
        &   & exp.&    \multicolumn{1}{c}{$jj44b$}&    \multicolumn{1}{c}{JUN45}&  \multicolumn{1}{c}{$fp$}\\\hline\hline
        $2^+_2$&    $0^+_1$&    59.5(1.4)&  6.1&    39.5&   12.0\\
         &          $2^+_1$&    40.5(1.2)&  93.9&   60.5&   88.0\\\hline
         $3^+_1$&   $2^+_1$&    9.6(11)&    34.7&    32.6&    94.1\\
         &          $2^+_2$&    81.8(5)&    65.3&    67.4&    4.5\\\hline
         $0^+_3$&   $2^+_1$&    $\leq3$&    99.9&    14.4&    8.6\\
         &          $2^+_2$&    100&    0.1& 85.4&    1.5 \\\hline
         $1^+_1$&   $0^+_1$&    59(4)&  0.0& 76.1&    76.4\\
         &          $2^+_1$&    14.7(15)&   81.0&    12.8&    23.5\\
         &          $0^+_2$&    9.8(11)&   0.1& 1.7&  0.0\\
         &          $2^+_2$&    16(6)&  18.8&    7.4& 0.1\\\hline
    \end{tabular}
    \label{tab:branchings}
    } 
\end{table}

The top panel of Fig.~\ref{fig:Is_M1} indicates that experiment and calculations in the $jj44$ model space agree on a sizable $M1$ scattering cross section around 3.5~MeV. It is then of interest to attempt to associate this strength with the experimental $1^+$ level at 3346~keV and compare data with predictions by the two effective interactions used within this space. Although this $1^+$ level exhibits a complex decay pattern (Table \ref{tab:branchings}) which is not captured completely by any calculation, the JUN45 interaction predicts the strongest branches to levels which match the experimental ones. The strong ground-state branch is explained by proton excitations within the $1p_{3/2}$ and between the $1p_{3/2}$ and $1p_{1/2}$ shells. The other three decay paths are best described by the addition of transitions within the neutron $0f_{5/2}$ orbital, which dominate in the JUN45 calculations, even though the branch to the $0^+_2$ state remains underestimated. The latter $0f_{5/2}$  orbital, together with its proton counterpart, also plays a dominant role in the preferential decay of the $3^+_1$ state to the $2^+_2$ level over the $2^+_1$ one as they are active only in the $3^+_1\rightarrow2^+_2$  branch. Both interactions compute branching ratios of the same magnitude and underestimate the strength of this branch somewhat (82\% vs. 65 or 67\%). 

Finally, the two excited $0^+$ states observed in this experiment deserve attention. As indicated in the introduction, the possibility of these levels being band heads of sequences associated with shapes differing from the ground state should not be overlooked.  In this context, it is noteworthy that while the $0^+_2$ state can only decay to the $2^+_1$ level, the $0^+_3$ one displays strong preference for decay to the $2^+_2$ level, with only a weak branch of $\leq3$\% to the $2^+_1$ state. The calculations with the JUN45 effective interaction again predict the experimental situation, underestimating the strength to the $2^+_2$ level by only $\sim$15\%. This decay branch is mainly driven by transitions within the proton $1p_{3/2}$ shell and the neutron $0g_{9/2}$ one as well as transitions between the neutron $1p_{1/2}$ and $0f_{5/2}$ orbitals, while the neutron $1p_{1/2}$ and $0f_{5/2}$ transitions are still active in the weak decay branch to the $2^+_1$ state with the proton excitations not contributing significantly in this instance. The strong preference for the decay to the $2^+_2$ state implies the dominant role of proton $1p_{3/2}$ and neutron $0g_{9/2}$ orbitals. Given the importance of the shape-driving $0g_{9/2}$ shell, this $0^+_3$ state remains a candidate for association with deformation. Unfortunately, no excited states built on this level could be identified in the coincidence data to place this conjecture on stronger footing. 

The $0^+_2$ level at 1656~keV is likely not associated with a highly-deformed shape. The proposed $2^+$ excitation based on the $0^+$ prolate state in $^{64}$Ni is only located 286~keV higher in excitation energy \cite{Marginean2020}, while the ($2^+$) level in $^{68}$Zn is linked to the $0^+_2$ state under consideration by a 2061-keV transition. On the other hand, in both the JUN45 and $jj44b$ calculations, the decay of this level is dominated by the neutron $0g_{9/2}$ shell; e.g., the presence of these excitations is similar to that noted above for the $0^+_3$ state. The involvement of the shape-driving orbital is consistent with the previous finding in Ref. \cite{Koizumi2004} suggesting soft triaxiality for the $0^+_2$ level. The different decay paths observed for the two excited $0^+$ states might then imply that their shapes differ.

\subsection{Comparison with NRF measurements in $^{66}$Zn}

The isotope $^{66}$Zn was also studied recently using NRF over a wide range of excitation energies \cite{Savran2022, Schwengner2021}. The results for this nucleus display several similarities to those for $^{68}$Zn presented here. At low energies, the $M1$ strength in both isotopes is distributed among several states, some of which are rather weak. For $^{66}$Zn, the total sum of the scattering cross section for the $1^-$ states is 5019(94)~eV~b, and is 278(20)~eV~b for $1^+$ ones, including only levels with definite $J^\pi$ assignments \cite{Schwengner2021}. For the present data on $^{68}$Zn, these values amount to 4370(150)~eV~b and 192(18)~eV~b, respectively. Contributions from states with unresolved parity could, within the $1\sigma$ limit, add up to an additional 1097 and 641~eV~b to each sum, respectively. Considering only the states with established parity, the total cross section of $1^+$ states makes up only about 4\% of the total spin-1 cross section in $^{68}$Zn, just as was the case in $^{66}$Zn \cite{Schwengner2021}. Considering states with uncertain parity, the fraction can fall between 3 and 16\%. In any case, the result echoes the finding in Ref. \cite{Schwengner2021} that mid-shell nuclei are characterized by significantly more $E1$ than $M1$ strength, in contrast to nuclei along the $N = 28$ line, whose dipole strength is distributed more evenly between levels with positive and negative parity \cite{Schwengner2020, Wilhelmy2018, Shizuma2017, Krishichayan2015}.

Reference \cite{Savran2022} also compared the average ground-state feeding as a function of excitation energy (ground-state branching ratio) obtained from the $^{66}$Zn NRF data measured at HI$\gamma$S with the results of a statistical model calculation. The experimental branching ratios were found to be significantly higher than the expectations from the model, with the latter predicting on average about 60\% of the measured values. For the present work on $^{68}$Zn, similar estimates were made to approximate the average elastic vs. inelastic branching ratios (see insert in Fig.~\ref{fig:TotalProjClover}). From these estimates, it is apparent that at excitation energies up to about 7~MeV, the ground-state branching ratio is roughly 80\%, which is in line with those calculated for the $^{66}$Zn data. However, above this energy, inelastic decays become dominant for $^{68}$Zn, with the inelastic yield through the $2^+$ state reaching 50\% around 8~MeV and increasing to 150\% at 9~MeV. The increase is less pronounced in the $^{66}$Zn data, where the inelastic branch also becomes stronger around 8~MeV, but does not reach values greater than $\sim50\%$ at higher energies. The difference of the observations between the two nuclei could be due in part to an underestimate of the ground-state strength in $^{68}$Zn because of the high level density. The $^{66}$Zn data was processed to take the detector response into account, which likely provides a more accurate estimate of the total ground-state intensity. Monte-Carlo simulations estimate that unresolved ground-state transitions could contribute up to an additional 40\% of the strength estimated at high beam energies in the present work. It is also worth noting that, for transitions with large $B(M1)$ values, shell-model calculations for both $^{66}$Zn and $^{68}$Zn predict ground-state branching ratios of the order of 50-80\% in line with the experimental observations.

Finally, in Ref. \cite{Schwengner2021} the $^{66}$Zn data were interpreted with shell-model calculations in two model spaces: one with a $^{48}$Ca core and another with the $^{56}$Ni one; i.e. with the $jj44$ model space used in the present work.  In all cases, effective interactions differing from those considered above were used. Nevertheless, in the case of $^{66}$Zn, it was found that the strength of the high-energy $M1$ transitions was best described by the former model space. In other words, excitations from the proton $0f_{7/2}$ to the $0f_{5/2}$ and $1p_{3/2}$ orbitals were found to be necessary to describe the magnitude of the $M1$ strength observed in $^{66}$Zn, a finding very similar to that discussed in the present work for $^{68}$Zn.

\section{Conclusion}

The unique low-spin experimental probe of NRF was exploited to investigate the structure of $^{68}$Zn across a wide range of excitation energies, from the ground state up to the particle emission threshold. The experiment directly populated 158 excited states for which many properties regarding their structure and decay were inferred. This was achieved by measuring the fluorescent radiation emitted from these excited states using, for the first time, the newly-established Clover Array at HI$\gamma$S. The properties measured include spin-parity quantum numbers, cross section values, and branching ratios. Many excited states below $\sim4$~MeV were populated indirectly and studied through $\gamma-\gamma$ coincidence measurements carried out using photon beams at energies just below the particle-emission threshold, where the nuclear density of states is high. Therefore, a large number of lower-lying states were populated in the $\gamma$-ray cascades originating from these high-lying spin-1 levels. States up to spin $J = 4$ were identified in this manner, significantly extending the reach of angular momentum accessible with photon beams.

Information on the structure of $^{68}$Zn has been greatly expanded, allowing for comparisons to large-scale shell-model calculations in a regime of spin and excitation energy where they are seldom tested. Comparisons between data and calculations demonstrate the importance of the inclusion of the $0g_{9/2}$ shell in describing the low-energy $M1$ strength in $^{68}$Zn based on the agreement between experimental and calculated cross section values using the $jj44b$ and JUN45 effective interactions in the $jj44$ model space. These calculations also account for the distribution of $E1$ excitations with energy. However, at high excitation energy, the experimental strengths can only be described by the inclusion of the $0f_{7/2}$ orbital in the full $fp$ model space, as available particle-hole excitations in the $jj44$ model space become exhausted before an excitation energy of 6~MeV, thus implying that the $^{56}$Ni core is broken at high energy. The low-energy level structure of $^{68}$Zn, studied in the coincidence measurements, is best described by the $jj44$ model space as well, with a satisfactory account of the experimental energies, spin-parity values, and branching ratios. A close parallel was found between the results obtained in the present NRF experiment and those of  similar measurements in $^{66}$Zn \cite{Savran2022, Schwengner2021}.

The results pose a challenge to nuclear theory. A large number of orbitals are required to describe exhaustive low-spin data such as those reported here, and calculations incorporating all the orbitals relevant for the description of the $M1$ and $E1$ strengths are not yet feasible. In the case of $^{68}$Zn, the number of single-particle states required runs into computation limitations, and effective interactions have not been developed for the expanded model space. It is expected that $^{68}$Zn is not unique in this requirement -- rarely has such an extensive test of shell-model calculations been performed. Future explorations that cover a suitably large excitation energy range may reveal similar phenomena.

\section{Acknowledgments}
The authors acknowledge C. Iliadis for providing a code which implements a Bayesian method for multipole mixing ratio determinations and for several discussions on the same topic. The authors also thank the GSI target-making facility for the loan of the enriched $^{68}$Zn target, the Army Research Laboratory and the United States Naval Academy for the loan of clover detectors, and the operating team at HI$\gamma$S for providing excellent beam conditions for this experimental campaign. 

This work is supported by the U.S. Department of Energy, Office of Science, Office of Nuclear Physics, under Grants No. DE-FG02-97ER41041 (UNC), No. DE-SC0023010 (UNC), No. DE-FG02-97ER41033 (Duke) and No. DE-FG02-97ER41042 (NCSU), and by the U.S. National Science Foundation under Grant and No. PHY-2110365 (FRIB, MSU).

\bibliography{bib}

\end{document}

%% file: Results/levels_and_decays_edits.tex
3346.09(20)$^\textit{b}$	& $1^+$	& 3343.9(6)	& 70.8(14)	& 	& 0	& $0^+_1$\\
	& 	& 2266.5(4)	& 17.5(13)	& $-3.4^{+0.9}_{-1.6}$ & 1077	& $2^+_1$\\
    &   &   &   &   $0.00\pm0.10$ & & \\
	& 	& 1689.1(4)	& 11.7(10)	& 	& 1656	& $0^+_2$\\
3680.8(8)	& $1^\pm,2^-$	& 3680.7(8)	& 	& 	& 0	& $0^+_1$\\
3715.7(9)	& $1^-$	& 3715.6(9)	& 	& 	& 0	& $0^+_1$\\
4026.7(8)	& $1^-$	& 4026.6(8)	& 59.5(18)	& 	& 0	& $0^+_1$\\
	& 	& 2369.1(5)	& 21.3(18)	& 	& 1656	& $0^+_2$\\
	& 	& 2948.7(6)	& 19.2(15)	& $-1.9^{+0.6}_{-0.7}$ 	& 1077	& $2^+_1$\\
        &   &   &   &  $-0.20^{+0.12}_{-0.18}$  & & \\

4331.9(9)	& $1^-$	& 4331.8(9)	& 42.1(22)	& 	& 0	& $0^+_1$\\
	& 	& 2446.3(4)	& 57.9(22)	& $-1.8^{+0.5}_{-0.6}$ 	& 1883	& $2^+_2$\\
        &   &   &   &   $-0.24^{+0.13}_{-0.23}$ & & \\
4338.1(10)	& $1^-$	& 4337.9(10)	& 62.5(17)	& 	& 0	& $0^+_1$\\
	& 	& 3260.3(6)	& 22.0(13)	& $-5.1^{+1.4}_{-2.6}$	& 1077	& $2^+_1$\\
            &   &   &   &   $0.10\pm0.07$ & & \\

	& 	& 2455.0(5)	& 15.5(17)	& $-5.6^{+2.5}_{-4.4}$ & 1883	& $2^+_2$\\
            &   &   &   &   $0.23^{+0.19}_{-0.17}$	 & & \\

4462.4(11)	& $1^+$	& 4462.3(11)	& 	& 	& 0	& $0^+_1$\\
4500.3(9)	& $1^-$	& 4500.1(9)	& 	& 	& 0	& $0^+_1$\\
4534.8(12)	& $1^+$	& 4534.7(12)	& 26.0(33)	& 	& 0	& $0^+_1$\\
	& 	& 3456.8(7)	& 74.0(33)	& $-2.6^{+0.9}_{-1.6}$ & 1077	& $2^+_1$\\
            &   &   &   &   $-0.07^{+0.13}_{-0.16}$	 & & \\

4608.3(12)	& $1^+$	& 4608.1(12)	& 	& 	& 0	& $0^+_1$\\
4681.9(11)	& $1^-$	& 4681.7(11)	& 	& 	& 0	& $0^+_1$\\
4721.2(12)	& $1^+$	& 4721.0(12)	& 	& 	& 0	& $0^+_1$\\
4793.0(13)	& $1^+$	& 4792.8(13)	& 	& 	& 0	& $0^+_1$\\
4992.0(10)$^\textit{b}$	& $1^-$	& 4992.9(14)	& 60.0(14)	& 	& 0	& $0^+_1$\\
	& 	& 3916.1(8)	& 40.0(14)	& $-2.5^{+0.4}_{-0.5}$ 	& 1077	& $2^+_1$\\
            &   &   &   &   $-0.09\pm0.06$ & & \\

5014.7(14)	& $1^+$	& 5014.5(14)	& 58.7(16)	& 	& 0	& $0^+_1$\\
	& 	& 3358.8(7)	& 26.4(13)	& 	& 1656	& $0^+_2$\\
	& 	& 3132.1(7)	& 15.0(16)	& $-0.7^{+0.5}_{-1.1}$	& 1883	& $2^+_2$\\
5130.6(14)	& $1^+,2^+$	& 5130.4(14)	& 	& 	& 0	& $0^+_1$\\
5242.8(16)	& $1^-$	& 5242.6(16)	& 	& 	& 0	& $0^+_1$\\
5266.8(15)	& $1^+$	& 5266.6(15)	& 59(5)	& 	& 0	& $0^+_1$\\
	& 	& 4190.2(12)	& 41(5)	& $-0.7^{+0.4}_{-0.7}$	& 1077	& $2^+_1$\\
5298.0(4)$^\textit{b}$	& $1^-$	& 5293.3(13)	& 85.8(11)	& 	& 0	& $0^+_1$\\
	& 	& 4221.3(10)	& 14.2(11)	& $-4.4^{+1.4}_{-2.9}$	& 1077	& $2^+_1$\\
            &   &   &   &   $0.07^{+0.10}_{-0.11}$ & & \\

5439.2(16)	& $1^+$	& 5439.0(16)	& 	& 	& 0	& $0^+_1$\\
5482.3(17)	& $1^-$	& 5482.0(17)	& 	& 	& 0	& $0^+_1$\\
5501.2(17)	& $1^-$	& 5500.9(17)	& 	& 	& 0	& $0^+_1$\\
5543.8(17)	& $1^-$	& 5543.5(17)	& 	& 	& 0	& $0^+_1$\\
5601.8(17)	& $1^-$	& 5601.5(17)	& 56.8(14)	& 	& 0	& $0^+_1$\\
	& 	& 4524.9(10)	& 43.2(14)	& $0.05^{+0.04}_{-0.05}$	& 1077	& $2^+_1$\\
5669.6(18)	& $1^-$	& 5669.4(18)	& 66(6)	& 	& 0	& $0^+_1$\\
	& 	& 4590.2(13)	& 34(6)	& $-0.7^{+0.7}_{-2.4}$	& 1077	& $2^+_1$\\
5856.4(16)	& $1^-$	& 5856.1(16)	& 	& 	& 0	& $0^+_1$\\
5866.9(16)	& $1^-$	& 5866.6(16)	& 91.6(19)	& 	& 0	& $0^+_1$\\
	& 	& 3531.6(11)	& 8.4(19)	& $-0.7^{+0.6}_{-1.3}$	& 2338	& $2^+_3$\\
5898.0(19)	& $1^-$	& 5897.7(19)	& 	& 	& 0	& $0^+_1$\\
5980.7(19)	& $1^-$	& 5980.4(19)	& 87.5(10)	& 	& 0	& $0^+_1$\\
	& 	& 4903.9(12)	& 12.5(10)	& $-3.1^{+0.8}_{-1.2}$	& 1077	& $2^+_1$\\
            &   &   &   &   $-0.02^{+0.09}_{-0.10}$ & & \\
6001.3(19)	& $1^-$	& 6001.0(19)	& 44.9(21)	& 	& 0	& $0^+_1$\\
	& 	& 4119.0(9)	& 55.1(21)	& $0.11^{+0.05}_{-0.06}$	& 1883	& $2^+_2$\\
6031.9(17)	& $1^-$	& 6031.6(17)	& 	& 	& 0	& $0^+_1$\\
	& 	& 4954.7(20)	& 	& 	& 1077	& $2^+_1$\\
	& 	& 4149.2(10)	& 	& $-3.1^{+1.0}_{-1.7}$	& 1883	& $2^+_2$\\
            &   &   &   &   $-0.02^{+0.12}_{-0.15}$ & & \\

6057.7(20)	& $1^-$	& 6057.4(20)	& 56.4(32)	& 	& 0	& $0^+_1$\\
	& 	& 4397.0(10)	& 43.6(32)	& 	& 1656	& $0^+_2$\\
6124.1(20)	& $1^-$	& 6123.8(20)	& 	& 	& 0	& $0^+_1$\\
6161.3(21)	& $1^-$	& 6161.0(21)	& 	& 	& 0	& $0^+_1$\\
6191.8(21)	& $1^-$	& 6191.4(21)	& 	& 	& 0	& $0^+_1$\\
6215.2(21)	& $1^-$	& 6214.9(21)	& 62.3(17)	& 	& 0	& $0^+_1$\\
	& 	& 5139.5(13)	& 37.7(17)	& $-12^{+4}_{-14}$ & 1077	& $2^+_1$\\
            &   &   &   &   $0.21\pm0.05$	 & & \\

6243.3(22)	& $1^-,2^-$	& 6243.0(22)	& 	& 	& 0	& $0^+_1$\\
6273.4(21)	& $1^-$	& 6273.1(21)	& 	& 	& 0	& $0^+_1$\\
6341.3(22)	& $1^-$	& 6341.0(22)	& 	& 	& 0	& $0^+_1$\\
6358.2(22)	& $1^-$	& 6357.8(22)	& 90.1(17)	& 	& 0	& $0^+_1$\\
	& 	& 5282.3(15)	& 9.9(17)	& $-0.75^{+0.35}_{-0.52}$	& 1077	& $2^+_1$\\
6385.1(22)	& $1^-$	& 6384.8(22)	& 	& 	& 0	& $0^+_1$\\
6404.8(22)	& $1^-$	& 6404.5(22)	& 	& 	& 0	& $0^+_1$\\
	& 	& 5329.7(23)	& 	& 	& 1077	& $2^+_1$\\
6486.6(23)	& $1^-$	& 6486.3(23)	& 	& 	& 0	& $0^+_1$\\
	& 	& 5410.0(23)	& 	& 	& 1077	& $2^+_1$\\
6506.0(23)	& $1^-$	& 6505.7(23)	& 	& 	& 0	& $0^+_1$\\
6564.9(23)	& $1^-$	& 6564.6(23)	& 	& 	& 0	& $0^+_1$\\
	& 	& 5487.9(24)	& 	& 	& 1077	& $2^+_1$\\
	& 	& 5487.1(24)	& 	& 	& 1077	& $2^+_1$\\
6585.9(23)	& $1^-$	& 6585.6(23)	& 	& 	& 0	& $0^+_1$\\
6633.3(24)	& $1^-$	& 6632.9(24)	& 	& 	& 0	& $0^+_1$\\
	& 	& 5557.2(24)	& 	& 	& 1077	& $2^+_1$\\
6697.8(24)	& $1^-$	& 6697.5(24)	& 	& 	& 0	& $0^+_1$\\
	& 	& 5623.2(24)	& 	& 	& 1077	& $2^+_1$\\
6746.9(25)	& $1^-$	& 6746.5(25)	& 	& 	& 0	& $0^+_1$\\
	& 	& 5669.4(27)	& 	& 	& 1077	& $2^+_1$\\
6799.1(25)	& $1^-$	& 6798.8(25)	& 	& 	& 0	& $0^+_1$\\
6841.5(25)	& $1^-$	& 6841.1(25)	& 	& 	& 0	& $0^+_1$\\
	& 	& 5766.0(26)	& 	& 	& 1077	& $2^+_1$\\
6860.4(25)	& $1^-$	& 6860.0(25)	& 	& 	& 0	& $0^+_1$\\
	& 	& 5783.9(25)	& 	& 	& 1077	& $2^+_1$\\
6874.4(28)	& $1^-$	& 6874.1(28)	& 	& 	& 0	& $0^+_1$\\
6929.9(26)	& $1^-$	& 6929.6(26)	& 	& 	& 0	& $0^+_1$\\
7037.2(26)	& $1^-$	& 7036.8(26)	& 82.7(25)	& 	& 0	& $0^+_1$\\
	& 	& 5389.6(15)	& 17.3(25)	& 	& 1656	& $0^+_2$\\
7102.9(23)	& $1^{(-)}$	& 7102.5(23)	& 	& 	& 0	& $0^+_1$\\
7132.4(27)	& $1^-$	& 7132.0(27)	& 80.7(31)	& 	& 0	& $0^+_1$\\
	& 	& 5251.1(23)	& 19.3(31)	& $-0.7^{+0.6}_{-1.6}$	& 1883	& $2^+_2$\\
7149.2(27)	& $1^-,2^-$	& 7148.8(27)	& 45.5(27)	& 	& 0	& $0^+_1$\\
	& 	& 5497.1(17)	& 28.9(25)	& 	& 1656	& $0^+_2$\\
	& 	& 5271.9(14)	& 25.6(24)	& $-5.6^{+2.2}_{-4.7}$	& 1883	& $2^+_2$\\
            &   &   &   &   $0.14^{+0.11}_{-0.12}$ & & \\
7202.6(28)	& $1^-$	& 7202.2(28)	& 	& 	& 0	& $0^+_1$\\
7346.2(30)	& $1^-$	& 7345.8(30)	& 	& 	& 0	& $0^+_1$\\
7359.1(30)	& $1^-$	& 7358.7(30)	& 	& 	& 0	& $0^+_1$\\
7397.9(30)	& $1^-$	& 7397.5(30)	& 	& 	& 0	& $0^+_1$\\
7414.9(30)	& $1^-$	& 7414.5(30)	& 	& 	& 0	& $0^+_1$\\
7424.1(30)	& $1^-$	& 7423.6(30)	& 	& 	& 0	& $0^+_1$\\
7433.7(30)	& $1^-$	& 7433.3(30)	& 56.3(26)	& 	& 0	& $0^+_1$\\
	& 	& 5782.4(22)	& 26.5(23)	& 	& 1656	& $0^+_2$\\
	& 	& 5092.4(18)	& 17.2(22)	& $-3.3^{+1.3}_{-2.4}$	& 2338	& $2^+_3$\\
            &   &   &   &   $0.01^{+0.16}_{-0.22}$ & & \\

7476.3(30)	& $1^-$	& 7475.9(30)	& 	& 	& 0	& $0^+_1$\\
7487.5(31)	& $1^-$	& 7487.0(31)	& 	& 	& 0	& $0^+_1$\\
7524(4)	& $1^-,2^+$	& 7523(4)	& 	& 	& 0	& $0^+_1$\\
7526.6(31)	& $1^-$	& 7526.2(31)	& 	& 	& 0	& $0^+_1$\\
7538.5(31)	& $1^-$	& 7538.1(31)	& 	& 	& 0	& $0^+_1$\\
7551.6(31)	& $1^-$	& 7551.1(31)	& 	& 	& 0	& $0^+_1$\\
7586.5(31)	& $1^-$	& 7586.0(31)	& 	& 	& 0	& $0^+_1$\\
7617.5(28)	& $1^-$	& 7617.1(28)	& 	& 	& 0	& $0^+_1$\\
7629.5(33)	& $1^\pm,2^-$	& 7629.0(33)	& 	& 	& 0	& $0^+_1$\\
7657.1(32)	& $1^-$	& 7656.7(32)	& 	& 	& 0	& $0^+_1$\\
7666.0(31)	& $1^-$	& 7665.5(31)	& 	& 	& 0	& $0^+_1$\\
7700.2(31)	& $1^-$	& 7699.7(31)	&  	& 	& 0	& $0^+_1$\\
	& 	& 6044.0(24)	& 	& 	& 1656	& $0^+_2$\\
7728.7(29)	& $1^\pm,2^-$	& 7728.3(29)	& 	& 	& 0	& $0^+_1$\\
	& 	& 6070.1(21)	& 	& 	& 1656	& $0^+_2$\\
7755.8(35)	& $1^-,2^-$	& 7755.3(35)	& 	& 	& 0	& $0^+_1$\\
7759.0(34)	& $1^-$	& 7758.5(34)	& 	& 	& 0	& $0^+_1$\\
7776(5)	& $1^\pm,2^-$	& 7775(5)	& 	& 	& 0	& $0^+_1$\\
7811.8(32)	& $1^-$	& 7811.3(32)	& 	& 	& 0	& $0^+_1$\\
7822.0(32)	& $1^-,2^-$	& 7821.5(32)	& 	& 	& 0	& $0^+_1$\\
7843(5)	& $1^-,2^-$	& 7842(5)	& 	& 	& 0	& $0^+_1$\\
7856.0(33)	& $1^-$	& 7855.5(33)	& 	& 	& 0	& $0^+_1$\\
7869.3(35)	& $1^-$	& 7868.8(35)	& 	& 	& 0	& $0^+_1$\\
7899.9(34)	& $1^{(-)}$	& 7899.4(34)	& 	& 	& 0	& $0^+_1$\\
7923.5(33)	& $1^-$	& 7923.0(33)	& 	& 	& 0	& $0^+_1$\\
7946.4(33)	& $1^-$	& 7945.9(33)	& 	& 	& 0	& $0^+_1$\\
7981.6(33)	& $1^-$	& 7981.1(33)	& 	& 	& 0	& $0^+_1$\\
7997.6(30)	& $1^{(-)}$	& 7997.1(30)	& 	& 	& 0	& $0^+_1$\\
8008.2(32)	& $1^\pm,2^+$	& 8007.7(32)	& 	& 	& 0	& $0^+_1$\\
8043.5(34)	& $1^-$	& 8043.0(34)	& 	& 	& 0	& $0^+_1$\\
8055.2(31)	& $1^{(-)}$	& 8054.7(31)	& 	& 	& 0	& $0^+_1$\\
	& 	& 5713.1(30)	& 	& $1.2^{+3.5}_{-0.7}$	& 2338	& $2^+_3$\\
8071.6(31)	& $1^-$	& 8071.1(30)	& 	& 	& 0	& $0^+_1$\\
	& 	& 5740.9(29)	& 	& $0.5^{+2.9}_{-0.5}$	& 2338	& $2^+_3$\\
	& 	& 6416.2(29)	& 	& 	& 1656	& $0^+_2$\\
8094.5(31)	& $1^{(-)}$	& 8094.0(31)	& 	& 	& 0	& $0^+_1$\\
8128(4)	& $1^-$	& 8128(4)	& 	& 	& 0	& $0^+_1$\\
	& 	& 5791(6)	& 	& $0.9^{+1.7}_{-0.5}$	& 2338	& $2^+_3$\\
8144(4)	& $1^-$	& 8143(4)	& 	& 	& 0	& $0^+_1$\\
	& 	& 6488.7(23)	& 	& 	& 1656	& $0^+_2$\\
8193.9(31)	& $1^-$	& 8193.3(31)	& 	& 	& 0	& $0^+_1$\\
8211.7(32)	& $1^-$	& 8211.2(32)	& 	& 	& 0	& $0^+_1$\\
8228(4)	& $1^-$	& 8228(4)	& 	& 	& 0	& $0^+_1$\\
8254(4)	& $1^{(-)}$	& 8253(4)	& 	& 	& 0	& $0^+_1$\\
	& 	& 6595.8(25)	& 	& 	& 1656	& $0^+_2$\\
8284(6)$^\textit{c}$	& $1^-$	& 8282(6)	& 	& 	& 0	& $0^+_1$\\
	& 	& 6627.7(30)	& 	& 	& 1656	& $0^+_2$\\
8305(6)$^\textit{c}$	& $1^-$	& 8303(6)	& 	& 	& 0	& $0^+_1$\\
8336.6(32)	& $1^\pm,2^-$	& 8336.0(32)	& 	& 	& 0	& $0^+_1$\\
8410(4)	& $1^-,2^+$	& 8409(4)	& 	& 	& 0	& $0^+_1$\\
8469.6(27)	& $1^-,2^-$	& 8469.0(27)	& 	& 	& 0	& $0^+_1$\\
8477(4)	& $1^{(-)}$	& 8476(4)	& 	& 	& 0	& $0^+_1$\\
8499.7(27)	& $1^{(-)}$	& 8499.1(27)	& 	& 	& 0	& $0^+_1$\\
8509.7(28)	& $1^{(-)}$	& 8509.2(28)	& 	& 	& 0	& $0^+_1$\\
8531(4)	& $1^{(-)}$	& 8530(4)	& 	& 	& 0	& $0^+_1$\\
8534.9(27)	& $1^\pm,2^-$	& 8534.3(27)	& 	& 	& 0	& $0^+_1$\\
8546.6(33)	& $1^-,2^-$	& 8546.0(33)	& 	& 	& 0	& $0^+_1$\\
8565.3(29)	& $1^-$	& 8564.7(29)	& 	& 	& 0	& $0^+_1$\\
	& 	& 5740(4)	& 	& $-0.5^{+0.6}_{-2.8}$	& 2822	& $2^+_4$\\
8587(5)	& $1^+$	& 8586(5)	& 	& 	& 0	& $0^+_1$\\
8602(4)	& $1^\pm,2^+$	& 8601(4)	& 	& 	& 0	& $0^+_1$\\
8606.2(28)	& $1^\pm,2^-$	& 8605.7(28)	& 	& 	& 0	& $0^+_1$\\
8627.4(28)	& $1^\pm,2^-$	& 8626.8(28)	& 	& 	& 0	& $0^+_1$\\
8715.2(32)	& $1^-,2^-$	& 8714.6(32)	& 	& 	& 0	& $0^+_1$\\
8752.8(31)	& $1^-$	& 8752.2(31)	& 	& 	& 0	& $0^+_1$\\
8798.6(31)	& $1^-$	& 8798.0(31)	& 	& 	& 0	& $0^+_1$\\
8821.6(29)	& $1^\pm,2^-$	& 8821.0(29)	& 	& 	& 0	& $0^+_1$\\
8844.0(29)	& $1^\pm,2^-$	& 8843.4(29)	& 	& 	& 0	& $0^+_1$\\
8881(4)	& $1^-,2^-$	& 8880(4)	& 	& 	& 0	& $0^+_1$\\
8930.6(31)	& $1^-,2^-$	& 8929.9(31)	& 	& 	& 0	& $0^+_1$\\
8955.9(31)	& $1^-$	& 8955.3(31)	& 	& 	& 0	& $0^+_1$\\
9011.4(32)	& $1^-,2^-$	& 9010.8(32)	& 	& 	& 0	& $0^+_1$\\
9039(4)	& $1^\pm,2^-$	& 9038(4)	& 	& 	& 0	& $0^+_1$\\
9092.7(33)	& $1^-,2^-$	& 9092.0(33)	& 	& 	& 0	& $0^+_1$\\
9106.5(33)	& $1^-,2^-$	& 9105.8(33)	& 	& 	& 0	& $0^+_1$\\
9133.2(31)	& $1^\pm,2^-$	& 9132.6(31)	& 	& 	& 0	& $0^+_1$\\
9154(5)	& $1,2^\pm$	& 9153(5)	& 	& 	& 0	& $0^+_1$\\
9185.1(31)	& $1^\pm,2^-$	& 9184.4(31)	& 	& 	& 0	& $0^+_1$\\
9200.1(34)	& $1,2^\pm$	& 9199.4(34)	& 	& 	& 0	& $0^+_1$\\
9246.9(31)	& $1^\pm,2^-$	& 9246.3(31)	& 	& 	& 0	& $0^+_1$\\
9257.8(34)	& $1^-$	& 9257.1(34)	& 	& 	& 0	& $0^+_1$\\
9265.5(33)	& $1^-$	& 9264.8(33)	& 	& 	& 0	& $0^+_1$\\
9274.9(34)	& $1^-$	& 9274.2(34)	& 	& 	& 0	& $0^+_1$\\
9290.0(35)	& $1^-,2^-$	& 9289.3(35)	& 	& 	& 0	& $0^+_1$\\
9315.3(35)	& $1^\pm,2^-$	& 9314.6(35)	& 	& 	& 0	& $0^+_1$\\
9359.4(35)	& $1^\pm,2^+$	& 9358.7(35)	& 	& 	& 0	& $0^+_1$\\
9383.5(34)	& $1^-$	& 9382.8(34)	& 	& 	& 0	& $0^+_1$\\
9393.5(35)	& $1^-$	& 9392.8(35)	& 	& 	& 0	& $0^+_1$\\
9435.7(34)	& $1^-$	& 9435.0(34)	& 	& 	& 0	& $0^+_1$\\
9444.2(34)	& $1^-$	& 9443.5(34)	& 	& 	& 0	& $0^+_1$\\
9490(14)	& $1^-,2^-$	& 9490(14)	& 	& 	& 0	& $0^+_1$\\
9523(6)	& $1^{(-)}$	& 9523(6)	& 	& 	& 0	& $0^+_1$\\
9553.5(35)	& $1^\pm,2^-$	& 9552.8(35)	& 	& 	& 0	& $0^+_1$\\
9590.8(35)	& $1^-$	& 9590.0(35)	& 	& 	& 0	& $0^+_1$\\
9626.2(35)	& $1^-,2^-$	& 9625.5(35)	& 	& 	& 0	& $0^+_1$\\
9685(4)	& $1^-,2^-$	& 9685(4)	& 	& 	& 0	& $0^+_1$\\
9702(4)	& $1^-,2^-$	& 9702(4)	& 	& 	& 0	& $0^+_1$\\
9797(4)	& $1^-$	& 9797(4)	& 	& 	& 0	& $0^+_1$\\
9864(6)	& $1^-$	& 9864(6)	& 	& 	& 0	& $0^+_1$\\
9892(4)	& $1^-$	& 9891(4)	& 	& 	& 0	& $0^+_1$\\
9907(4)	& $1^-,2^-$	& 9907(4)	& 	& 	& 0	& $0^+_1$\\
9928(7)	& $1^-$	& 9927(7)	& 	& 	& 0	& $0^+_1$

%% file: Results/CrossSectionsEditedEnergies.tex
3346&$1^+$&30(5)$^\textit{a}$\\
3681&$1$&3.2(9), 2.4(6)\\
3716&$1^-$&7.2(19)$^\textit{a}$\\
4027&$1^-$&19(5)\\
4332&$1^-$&4.4(11)\\
4338&$1^-$&23(6)$^\textit{a}$\\
4462&$1^+$&38(11)$^\textit{a}$\\
4500&$1^-$&22(7)$^\textit{a}$\\
4535&$1^+$&1.6(4)\\
4608&$1^+$&46(11)\\
4682&$1^-$&62(15)\\
4721&$1^+$&10.1(2.4)\\
4793&$1^+$&17(4)\\
4992&$1^-$&28(6)\\
5015&$1^+$&18(4)\\
5131&$1^+$&8.8(22)\\
5243&$1^-$&3.8(10)\\
5267&$1^+$&4.4(11)\\
5298&$1^-$&53(12)\\
5439&$1^+$&5.7(14)\\
5482&$1^-$&9.4(21)\\
5501&$1^-$&73(16)\\
5544&$1^-$&12.1(27)\\
5602&$1^-$&26(6)\\
5670&$1^-$&8.7(22)\\
5856&$1^-$&--\\
5867&$1^-$&--\\
5898&$1^-$&--\\
5981&$1^-$&94(22)\\
6001&$1^-$&17(4)\\
6032&$1^-$&69(16)\\
6058&$1^-$&10.7(26)\\
6124&$1^-$&20(5)\\
6161&$1^-$&40(9)\\
6192&$1^-$&31(7)\\
6215&$1^-$&52(13)\\
6243&$1^-$&8.9(24)\\
6273&$1^-$&53(13)\\
6341&$1^-$&37(8)\\
6358&$1^-$&81(18)\\
6385&$1^-$&20(4)\\
6405&$1^-$&62(14)\\
6487&$1^-$&41(10)\\
6506&$1^-$&20(5)\\
6565&$1^-$&37(9)\\
6586&$1^-$&49(12)\\
6633&$1^-$&107(25)\\
6698&$1^-$&54(13)\\
6747&$1^-$&20(5)\\
6799&$1^-$&67(15)\\
6842&$1^-$&37(9)\\
6860&$1^-$&67(16)\\
6874&$1^-$&24(6)\\
6930&$1^-$&68(16)\\
7037&$1^-$&43(10)\\
7103&$1$&39(9), 18(4)\\
7132&$1^-$&31(7)\\
7149&$1^-$&21(5)\\
7203&$1^-$&54(13)\\
7346&$1^-$&38(10)\\
7359&$1^-$&24(6)\\
7398&$1^-$&48(11)\\
7415&$1^-$&55(13)\\
7424&$1^-$&41(10)\\
7434&$1^-$&38(9)\\
7476&$1^-$&44(11)\\
7488&$1^-$&25(6)\\
7524&$1^-$&10.6(28)\\
7527&$1^-$&35(9)\\
7539&$1^-$&29(7)\\
7552&$1^-$&87(19)\\
7587&$1^-$&36(8)\\
7618&$1^-$&47(11)\\
7630&$1$&23(6), 9.1(24)\\
7657&$1^-$&31(8)\\
7666&$1^-$&49(12)\\
7700&$1^-$&77(18)\\
7729&$1$&16(4), 11.3(28)\\
7756&$1^-$&38(10)\\
7759&$1^-$&50(12)\\
7776&$1$&6.5(22), 4.8(14)\\
7812&$1^-$&38(9)\\
7822&$1^-$&30(7)\\
7843&$1^-$&8.4(2.7)\\
7856&$1^-$&26(6)\\
7869&$1^-$&14(4)\\
7900&$1$&22(6), 8.8(25)\\
7924&$1^-$&46(11)\\
7946&$1^-$&22(6)\\
7982&$1^-$&70(17)\\
7998&$1$&45(11), 16(4)\\
8008&$1$&11.5(32), 11.0(28)\\
8044&$1^-$&43(10)\\
8055&$1$&37(9), 15(4)\\
8072&$1^-$&64(15)\\
8095&$1$&28(7), 16(4)\\
8128&$1^-$&48(12)\\
8144&$1^-$&31(9)\\
8194&$1^-$&79(19)\\
8212&$1^-$&35(9)\\
8228&$1^-$&23(6)\\
8254&$1$&14(4), 8.6(26)\\
8284&$1^-$&143(34)\\
8305&$1^-$&134(32)\\
8337&$1$&48(11), 49(11)\\
8410&$1^-$&23(7)\\
8470&$1^-$&32(8)\\
8477&$1$&19(6), 10.0(30)\\
8500&$1$&44(11), 18(5)\\
8510&$1$&33(8), 15(4)\\
8531&$1$&33(9), 19(5)\\
8535&$1$&38(10), 25(6)\\
8547&$1^-$&23(6)\\
8565&$1^-$&26(7)\\
8587&$1^+$&13.0(35)\\
8602&$1$&19(6), 18(5)\\
8606&$1$&31(8), 24(6)\\
8627&$1$&19(5), 21(5)\\
8715&$1^-$&30(8)\\
8753&$1^-$&38(10)\\
8799&$1^-$&85(20)\\
8822&$1$&46(11), 25(6)\\
8844&$1$&28(7), 25(6)\\
8881&$1^-$&13(4)\\
8931&$1^-$&30(7)\\
8956&$1^-$&70(17)\\
9011&$1^-$&22(6)\\
9039&$1$&23(7), 14(4)\\
9093&$1^-$&27(7)\\
9107&$1^-$&27(7)\\
9133&$1$&25(6), 14(4)\\
9154&$1$&16(4), 8.7(24)\\
9185&$1$&17(5), 15(4)\\
9200&$1$&15(4), 8.1(23)\\
9247&$1$&20(6), 14(4)\\
9258&$1$&33(10), 10(4)\\
9266&$1$&50(13), 12(4)\\
9275&$1^-$&31(9)\\
9290&$1^-$&26(7)\\
9315&$1$&18(5), 12(4)\\
9359&$1$&20(5), 13(4)\\
9384&$1^-$&18(5)\\
9394&$1^-$&18(5)\\
9436&$1^-$&29(8)\\
9444&$1^-$&28(7)\\
9490&$1^-$&9.2(31)\\
9523&$1$&16(4), 4.6(17)\\
9554&$1$&13(4), 10.8(29)\\
9591&$1^-$&18(5)\\
9626&$1^-$&14(5)\\
9685&$1^-$&30(8)\\
9702&$1^-$&27(7)\\
9797&$1^-$&32(8)\\
9864&$1^-$&23(6)\\
9892&$1^-$&61(15)\\
9907&$1^-$&29(8)\\
9928&$1^-$&21(7)

%% file: bib.bib
@article{Crider2016,
   author = {B. P. Crider and C. J. Prokop and S. N. Liddick and M. Al-Shudifat and A. D. Ayangeakaa and M. P. Carpenter and J. J. Carroll and J. Chen and C. J. Chiara and H. M. David and A. C. Dombos and S. Go and R. Grzywacz and J. Harker and R. V. F. Janssens and N. Larson and T. Lauritsen and R. Lewis and S. J. Quinn and F. Recchia and A. Spyrou and S. Suchyta and W. B. Walters and S. Zhu},
   doi = {10.1016/j.physletb.2016.10.020},
   issn = {03702693},
   journal = {Physics Letters B},
   month = {12},
   pages = {108-113},
   title = {Shape coexistence from lifetime and branching-ratio measurements in 68,70Ni},
   volume = {763},
   year = {2016}
}

@article{Leoni2017,
   author = {S. Leoni and B. Fornal and N. Mărginean and M. Sferrazza and Y. Tsunoda and T. Otsuka and G. Bocchi and F. C. L. Crespi and A. Bracco and S. Aydin and M. Boromiza and D. Bucurescu and N. Cieplicka-Oryǹczak and C. Costache and S. Călinescu and N. Florea and D. G. Ghiţă and T. Glodariu and A. Ionescu and Ł.W. Iskra and M. Krzysiek and R. Mărginean and C. Mihai and R. E. Mihai and A. Mitu and A. Negreţ and C. R. Niţă and A. Olăcel and A. Oprea and S. Pascu and P. Petkov and C. Petrone and G. Porzio and A. Şerban and C. Sotty and L. Stan and I. Ştiru and L. Stroe and R. Şuvăilă and S. Toma and A. Turturică and S. Ujeniuc and C. A. Ur},
   doi = {10.1103/PhysRevLett.118.162502},
   issn = {0031-9007},
   issue = {16},
   journal = {Physical Review Letters},
   month = {4},
   pages = {162502},
   title = {Multifaceted Quadruplet of Low-Lying Spin-Zero States in <math display="inline"> <mrow> <mmultiscripts> <mrow> <mi>Ni</mi> </mrow> <mprescripts/> <none/> <mrow> <mn>66</mn> </mrow> </mmultiscripts> </mrow> </math> : Emergence of Shape Isomerism in Light Nuclei},
   volume = {118},
   year = {2017}
}

@article{Prokop2015,
   author = {C. J. Prokop and B. P. Crider and S. N. Liddick and A. D. Ayangeakaa and M. P. Carpenter and J. J. Carroll and J. Chen and C. J. Chiara and H. M. David and A. C. Dombos and S. Go and J. Harker and R. V. F. Janssens and N. Larson and T. Lauritsen and R. Lewis and S. J. Quinn and F. Recchia and D. Seweryniak and A. Spyrou and S. Suchyta and W. B. Walters and S. Zhu},
   doi = {10.1103/PhysRevC.92.061302},
   issn = {0556-2813},
   issue = {6},
   journal = {Physical Review C},
   month = {12},
   pages = {061302},
   title = {New low-energy <math> <msup> <mn>0</mn> <mo>+</mo> </msup> </math> state and shape coexistence in <math> <mmultiscripts> <mi>Ni</mi> <mprescripts/> <none/> <mn>70</mn> </mmultiscripts> </math>},
   volume = {92},
   year = {2015}
}

@article{Marginean2020,
   author = {N. Mărginean and D. Little and Y. Tsunoda and S. Leoni and R. V. F. Janssens and B. Fornal and T. Otsuka and C. Michelagnoli and L. Stan and F. C. L. Crespi and C. Costache and R. Lica and M. Sferrazza and A. Turturica and A. D. Ayangeakaa and K. Auranen and M. Barani and P. C. Bender and S. Bottoni and M. Boromiza and A. Bracco and S. Călinescu and C. M. Campbell and M. P. Carpenter and P. Chowdhury and M. Ciemała and N. Cieplicka-Ory\`{n}czak and D. Cline and C. Clisu and H. L. Crawford and I. E. Dinescu and J. Dudouet and D. Filipescu and N. Florea and A. M. Forney and S. Fracassetti and A. Gade and I. Gheorghe and A. B. Hayes and I. Harca and J. Henderson and A. Ionescu and Ł. W. Iskra and M. Jentschel and F. Kandzia and Y. H. Kim and F. G. Kondev and G. Korschinek and U. Köster and Krishichayan and M. Krzysiek and T. Lauritsen and J. Li and R. Mărginean and E. A. Maugeri and C. Mihai and R. E. Mihai and A. Mitu and P. Mutti and A. Negret and C. R. Niţă and A. Olăcel and A. Oprea and S. Pascu and C. Petrone and C. Porzio and D. Rhodes and D. Seweryniak and D. Schumann and C. Sotty and S. M. Stolze and R. Şuvăilă and S. Toma and S. Ujeniuc and W. B. Walters and C. Y. Wu and J. Wu and S. Zhu and S. Ziliani},
   doi = {10.1103/PhysRevLett.125.102502},
   issn = {0031-9007},
   issue = {10},
   journal = {Physical Review Letters},
   month = {9},
   pages = {102502},
   title = {Shape Coexistence at Zero Spin in <math display="inline"> <mrow> <mmultiscripts> <mrow> <mi>Ni</mi> </mrow> <mprescripts/> <none/> <mrow> <mn>64</mn> </mrow> </mmultiscripts> </mrow> </math> Driven by the Monopole Tensor Interaction},
   volume = {125},
   year = {2020}
}

@article{Little2022,
   author = {D. Little and A. D. Ayangeakaa and R. V. F. Janssens and S. Zhu and Y. Tsunoda and T. Otsuka and B. A. Brown and M. P. Carpenter and A. Gade and D. Rhodes and C. R. Hoffman and F. G. Kondev and T. Lauritsen and D. Seweryniak and J. Wu and J. Henderson and C. Y. Wu and P. Chowdhury and P. C. Bender and A. M. Forney and W. B. Walters},
   doi = {10.1103/PhysRevC.106.044313},
   issn = {2469-9985},
   issue = {4},
   journal = {Physical Review C},
   month = {10},
   pages = {044313},
   title = {Multistep Coulomb excitation of <math> <mmultiscripts> <mi>Ni</mi> <mprescripts/> <none/> <mn>64</mn> </mmultiscripts> </math> : Shape coexistence and nature of low-spin excitations},
   volume = {106},
   year = {2022}
}

@article{Tsunoda2014,
   author = {Yusuke Tsunoda and Takaharu Otsuka and Noritaka Shimizu and Michio Honma and Yutaka Utsuno},
   doi = {10.1103/PhysRevC.89.031301},
   issn = {0556-2813},
   issue = {3},
   journal = {Physical Review C},
   month = {3},
   pages = {031301},
   title = {Novel shape evolution in exotic Ni isotopes and configuration-dependent shell structure},
   volume = {89},
   year = {2014}
}

@article{Olaizola2017,
   author = {B. Olaizola and L. M. Fraile and H. Mach and A. Poves and F. Nowacki and A. Aprahamian and J. A. Briz and J. Cal-González and D. Ghiţa and U. Köster and W. Kurcewicz and S. R. Lesher and D. Pauwels and E. Picado and D. Radulov and G. S. Simpson and J. M. Udías},
   doi = {10.1103/PhysRevC.95.061303},
   issn = {2469-9985},
   issue = {6},
   journal = {Physical Review C},
   month = {6},
   pages = {061303},
   title = {Search for shape-coexisting <math> <msup> <mn>0</mn> <mo>+</mo> </msup> </math> states in <math> <mmultiscripts> <mi>Ni</mi> <mprescripts/> <none/> <mn>66</mn> </mmultiscripts> </math> from lifetime measurements},
   volume = {95},
   year = {2017}
}

@article{Koizumi2004,
   author = {M Koizumi and A Seki and Y Toh and A Osa and Y Utsuno and A Kimura and M Oshima and T Hayakawa and Y Hatsukawa and J Katakura and M Matsuda and T Shizuma and T Czosnyka and M Sugawara and T Morikawa and H Kusakari},
   doi = {https://doi.org/10.1016/j.nuclphysa.2003.10.010},
   issn = {0375-9474},
   issue = {1},
   journal = {Nuclear Physics A},
   pages = {46-58},
   title = {Multiple Coulomb excitation experiment of 68Zn},
   volume = {730},
   url = {https://www.sciencedirect.com/science/article/pii/S0375947403017950},
   year = {2004}
}

@article{McCutchan2012,
   author = {E.A. McCutchan},
   doi = {10.1016/j.nds.2012.06.002},
   issn = {00903752},
   issue = {6-7},
   journal = {Nuclear Data Sheets},
   month = {6},
   pages = {1735-1870},
   title = {Nuclear Data Sheets for A = 68},
   volume = {113},
   year = {2012}
}

@article{Devlin1999,
   author = {M. Devlin and A. V. Afanasjev and R. M. Clark and D. R. LaFosse and I. Y. Lee and F. Lerma and A. O. Macchiavelli and R. W. MacLeod and I. Ragnarsson and P. Ring and D. Rudolph and D. G. Sarantites and P. G. Thirolf},
   doi = {10.1103/PhysRevLett.82.5217},
   issn = {0031-9007},
   issue = {26},
   journal = {Physical Review Letters},
   month = {6},
   pages = {5217-5220},
   title = {Superdeformation in <math display="inline"> <mrow> <msup> <mrow> <mi/> </mrow> <mrow> <mn>68</mn> </mrow> </msup> </mrow> <mi>Zn</mi> </math> : Evidence for a New, Neutron-Rich Island of Superdeformation in <math display="inline"> <mi mathvariant="italic">A</mi> <mi/> <mi>∼</mi> <mi/> <mn>70</mn> </math> Nuclei},
   volume = {82},
   year = {1999}
}

@article{Moreh1983,
   author = {R Moreh and O Shahal and J Tenenbaum},
   doi = {10.1088/0305-4616/9/7/011},
   issn = {0305-4616},
   issue = {7},
   journal = {Journal of Physics G: Nuclear Physics},
   month = {7},
   pages = {755-762},
   title = {Photoexcitation of levels at 6605 keV in <sup>48</sup> Ti and 7362 keV in <sup>68</sup> Zn},
   volume = {9},
   year = {1983}
}

@article{Savran2022,
   author = {D. Savran and J. Isaak and R. Schwengner and R. Massarczyk and M. Scheck and W. Tornow and G. Battaglia and T. Beck and S. W. Finch and C. Fransen and U. Friman-Gayer and R. Gonzalez and E. Hoemann and R. V. F. Janssens and S. R. Johnson and M. D. Jones and J. Kleemann and Krishichayan and D. R. Little and D. O'Donnell and O. Papst and N. Pietralla and J. Sinclair and V. Werner and O. Wieland and J. Wilhelmy},
   doi = {10.1103/PhysRevC.106.044324},
   issn = {2469-9985},
   issue = {4},
   journal = {Physical Review C},
   month = {10},
   pages = {044324},
   title = {Model-independent determination of the dipole response of <math> <mmultiscripts> <mi>Zn</mi> <mprescripts/> <none/> <mn>66</mn> </mmultiscripts> </math> using quasimonoenergetic and linearly polarized photon beams},
   volume = {106},
   year = {2022}
}

@article{Schwengner2021,
   author = {R. Schwengner and R. Massarczyk and M. Scheck and W. Tornow and G. Battaglia and T. Beck and D. Bemmerer and N. Benouaret and R. Beyer and M. Butterling and F. Fiedler and S. W. Finch and C. Fransen and U. Friman-Gayer and A. Frotscher and R. Gonzalez and M. Grieger and A. Hartmann and T. Hensel and E. Hoemann and H. Hoffmann and R. V. F. Janssens and S. Johnson and M. D. Jones and A. R. Junghans and N. Kelly and J. Kleemann and Krishichayan and D. R. Little and F. Ludwig and S. E. Müller and D. O'Donnell and O. Papst and E. Pirovano and J. Sinclair and M. P. Takács and S. Turkat and S. Urlaß and A. Wagner and V. Werner and O. Wieland and J. Wilhelmy},
   doi = {10.1103/PhysRevC.103.024312},
   issn = {2469-9985},
   issue = {2},
   journal = {Physical Review C},
   month = {2},
   pages = {024312},
   title = {Electric and magnetic dipole strength in $^\{66\}$Zn},
   volume = {103},
   year = {2021}
}

@article{Zilges2022,
   author = {A. Zilges and D.L. Balabanski and J. Isaak and N. Pietralla},
   doi = {10.1016/j.ppnp.2021.103903},
   issn = {01466410},
   journal = {Progress in Particle and Nuclear Physics},
   month = {1},
   pages = {103903},
   title = {Photonuclear reactions—From basic research to applications},
   volume = {122},
   year = {2022}
}

@article{Ayangeakaa2021,
   author = {A. D. Ayangeakaa and U. Friman-Gayer and R. V. F. Janssens},
   journal = {Innovation News Network},
   title = {The Clover Array for Nuclear Structure Studies at HIγS},
   year = {2021},
   note={https://www.innovationnewsnetwork.com/nuclear-structure/10491/}
}

@misc{FrimanPapst2022,
   author = {U. Friman-Gayer and O. Papst},
   title = {nutr: new utr},
   note = {https://github.com/u-eff-gee/nutr},
   year = {2022}
}

@article{Iliadis2021,
   author = {Christian Iliadis and Udo Friman-Gayer},
   doi = {10.1140/epja/s10050-021-00472-1},
   issn = {1434-6001},
   issue = {6},
   journal = {The European Physical Journal A},
   month = {6},
   pages = {190},
   title = {Linear polarization–direction correlations in $\gamma$-ray scattering experiments},
   volume = {57},
   year = {2021}
}

@phdthesis{Johnson2025,
   author = {S. R. Johnson},
   city = {Chapel Hill},
   school = {University of North Carolina at Chapel Hill},
   title = {From the ground state to the particle emission threshold: Nuclear resonance fluorescence in $^{68}\text{Zn}$},
   year = {2025}
}

@article{Otsuka2016,
   author = {T Otsuka and Y Tsunoda},
   doi = {10.1088/0954-3899/43/2/024009},
   issn = {0954-3899},
   issue = {2},
   journal = {Journal of Physics G: Nuclear and Particle Physics},
   month = {2},
   pages = {024009},
   title = {The role of shell evolution in shape coexistence},
   volume = {43},
   year = {2016}
}

@article{Otsuka2005,
   author = {Takaharu Otsuka and Toshio Suzuki and Rintaro Fujimoto and Hubert Grawe and Yoshinori Akaishi},
   doi = {10.1103/PhysRevLett.95.232502},
   issn = {0031-9007},
   issue = {23},
   journal = {Physical Review Letters},
   month = {11},
   pages = {232502},
   title = {Evolution of Nuclear Shells due to the Tensor Force},
   volume = {95},
   year = {2005}
}

@article{Johnson2023,
   author = {S. R. Johnson and R. V. F. Janssens and U. Friman-Gayer and B. A. Brown and B. P. Crider and S. W. Finch and Krishichayan and D. R. Little and S. Mukhopadhyay and E. E. Peters and A. P. D. Ramirez and J. A. Silano and A. P. Tonchev and W. Tornow and S. W. Yates},
   doi = {10.1103/PhysRevC.108.024315},
   issn = {2469-9985},
   issue = {2},
   journal = {Physical Review C},
   month = {8},
   pages = {024315},
   title = {Testing shell-model interactions at high excitation energy and low spin: Nuclear resonance fluorescence in <math> <mmultiscripts> <mi>Ge</mi> <mprescripts/> <none/> <mn>74</mn> </mmultiscripts> </math>},
   volume = {108},
   year = {2023}
}

@article{Weller2009,
   author = {Henry R. Weller and Mohammad W. Ahmed and Haiyan Gao and Werner Tornow and Ying K. Wu and Moshe Gai and Rory Miskimen},
   doi = {10.1016/j.ppnp.2008.07.001},
   issn = {01466410},
   issue = {1},
   journal = {Progress in Particle and Nuclear Physics},
   month = {1},
   pages = {257-303},
   title = {Research opportunities at the upgraded HIγS facility},
   volume = {62},
   url = {https://linkinghub.elsevier.com/retrieve/pii/S0146641008000434},
   year = {2009}
}

@article{Metzger1972,
   author = {F.R. Metzger},
   doi = {10.1016/0375-9474(72)90304-1},
   issn = {03759474},
   issue = {2},
   journal = {Nuclear Physics A},
   month = {7},
   pages = {409-416},
   title = {Radiative widths of spin-1 levels in the Zn isotopes},
   volume = {189},
   year = {1972}
}

@article{Honma2005,
   author = {M. Honma and T. Otsuka and B. A. Brown and T. Mizusaki},
   doi = {10.1140/epjad/i2005-06-032-2},
   issn = {1434-6001},
   issue = {S1},
   journal = {The European Physical Journal A},
   month = {9},
   pages = {499-502},
   title = {Shell-model description of neutron-rich pf-shell nuclei with a new effective interaction GXPF 1},
   volume = {25},
   year = {2005}
}

@article{Broda2012,
   author = {R. Broda and T. Pawłat and W. Królas and R. V. F. Janssens and S. Zhu and W. B. Walters and B. Fornal and C. J. Chiara and M. P. Carpenter and N. Hoteling and Ł. W. Iskra and F. G. Kondev and T. Lauritsen and D. Seweryniak and I. Stefanescu and X. Wang and J. Wrzesiński},
   doi = {10.1103/PhysRevC.86.064312},
   issn = {0556-2813},
   issue = {6},
   journal = {Physical Review C},
   month = {12},
   pages = {064312},
   title = {Spectroscopic study of the <math display="inline"> <msup> <mrow/> <mrow> <mn>64</mn> <mo>,</mo> <mn>66</mn> <mo>,</mo> <mn>68</mn> </mrow> </msup> </math> Ni isotopes populated in <math display="inline"> <msup> <mrow/> <mn>64</mn> </msup> </math> Ni +  <math display="inline"> <msup> <mrow/> <mn>238</mn> </msup> </math> U collisions},
   volume = {86},
   year = {2012}
}

@article{Suchyta2014,
   author = {S. Suchyta and S. N. Liddick and Y. Tsunoda and T. Otsuka and M. B. Bennett and A. Chemey and M. Honma and N. Larson and C. J. Prokop and S. J. Quinn and N. Shimizu and A. Simon and A. Spyrou and V. Tripathi and Y. Utsuno and J. M. VonMoss},
   doi = {10.1103/PhysRevC.89.021301},
   issn = {0556-2813},
   issue = {2},
   journal = {Physical Review C},
   month = {2},
   pages = {021301},
   title = {Shape coexistence in <math> <mmultiscripts> <mi>Ni</mi> <mprescripts/> <none/> <mn>68</mn> </mmultiscripts> </math>},
   volume = {89},
   year = {2014}
}

@article{Junde2011,
   author = {Huo Junde and Huo Su and Yang Dong},
   doi = {10.1016/j.nds.2011.04.004},
   issn = {00903752},
   issue = {6},
   journal = {Nuclear Data Sheets},
   month = {6},
   pages = {1513-1645},
   title = {Nuclear Data Sheets for A = 56},
   volume = {112},
   year = {2011}
}

@article{Ouellet2011,
   author = {Christian Ouellet and Balraj Singh},
   doi = {10.1016/j.nds.2011.08.004},
   issn = {00903752},
   issue = {9},
   journal = {Nuclear Data Sheets},
   month = {9},
   pages = {2199-2355},
   title = {Nuclear Data Sheets for A = 32},
   volume = {112},
   year = {2011}
}

@misc{Dudouet2024,
   author = {Jérémie Dudouet},
   title = {Cubix},
   note = {https://doi.org/10.5281/zenodo.10683242},
   year = {2024}
}

@article{Brown2014,
   author = {B. A. Brown and W. D. M. Rae},
   doi = {10.1016/j.nds.2014.07.022},
   issn = {00903752},
   journal = {Nuclear Data Sheets},
   month = {6},
   pages = {115-118},
   title = {The Shell-Model Code NuShellX@MSU},
   volume = {120},
   year = {2014}
}

@article{Honma2009,
   author = {M. Honma and T. Otsuka and T. Mizusaki and M. Hjorth-Jensen},
   doi = {10.1103/PhysRevC.80.064323},
   issn = {0556-2813},
   issue = {6},
   journal = {Physical Review C},
   month = {12},
   pages = {064323},
   title = {New effective interaction for <math display="inline"> <mrow> <msub> <mi>f</mi> <mrow> <mn>5</mn> </mrow> </msub> <msub> <mi mathvariant="italic">pg</mi> <mrow> <mn>9</mn> </mrow> </msub> </mrow> </math> -shell nuclei},
   volume = {80},
   year = {2009}
}

@article{Mukhopadhyay2017,
   author = {S. Mukhopadhyay and B. P. Crider and B. A. Brown and S. F. Ashley and A. Chakraborty and A. Kumar and M. T. McEllistrem and E. E. Peters and F. M. Prados-Estévez and S. W. Yates},
   doi = {10.1103/PhysRevC.95.014327},
   issn = {2469-9985},
   issue = {1},
   journal = {Physical Review C},
   month = {1},
   pages = {014327},
   title = {Nuclear structure of   Ge   76   from inelastic neutron scattering measurements and shell model calculations},
   volume = {95},
   url = {https://link.aps.org/doi/10.1103/PhysRevC.95.014327},
   year = {2017}
}

@article{Kelly2026,
  title = {Detailed View at Magnetic Dipole Strengths: The Case of Semimagic ${}^{50}\mathrm{Ti}$},
  author = {Kelly, B. and Spieker, M. and Friman-Gayer, U. and Baby, L. T. and Beck, T. and Conley, A. L. and Finch, S. W. and Isaak, J. and Krishichayan and Litvinova, E. and Pai, H. and Pietralla, N. and Savran, D. and Tornow, W. and Tsoneva, N. and Volya, A. and Werner, V.},
  journal = {Phys. Rev. Lett.},
  volume = {136},
  issue = {8},
  pages = {082502},
  numpages = {7},
  year = {2026},
  month = {Feb},
  publisher = {American Physical Society},
  doi = {10.1103/82y9-svrd},
  url = {https://link.aps.org/doi/10.1103/82y9-svrd}
}

@article{Dinca2005,
   author = {D.-C. Dinca and R. V. F. Janssens and A. Gade and D. Bazin and R. Broda and B. A. Brown and C. M. Campbell and M. P. Carpenter and P. Chowdhury and J. M. Cook and A. N. Deacon and B. Fornal and S. J. Freeman and T. Glasmacher and M. Honma and F. G. Kondev and J.-L. Lecouey and S. N. Liddick and P. F. Mantica and W. F. Mueller and H. Olliver and T. Otsuka and J. R. Terry and B. A. Tomlin and K. Yoneda},
   doi = {10.1103/PhysRevC.71.041302},
   issn = {0556-2813},
   issue = {4},
   journal = {Physical Review C},
   month = {4},
   pages = {041302},
   title = {Reduced transition probabilities to the first $2^+$ state in $^\{52,54,56\}$Ti and development of shell closures at N=32,34},
   volume = {71},
   year = {2005}
}

@article{Huck1985,
   author = {A. Huck and G. Klotz and A. Knipper and C. Miehé and C. Richard-Serre and G. Walter and A. Poves and H. L. Ravn and G. Marguier},
   doi = {10.1103/PhysRevC.31.2226},
   issn = {0556-2813},
   issue = {6},
   journal = {Physical Review C},
   month = {6},
   pages = {2226-2237},
   title = {Beta decay of the new isotopes $^\{52\}$K, $^\{52\}$Ca, and $^\{52\}$Sc; a test of the shell model far from stability},
   volume = {31},
   year = {1985}
}

@article{Janssens2002,
   author = {R. V. F. Janssens and B. Fornal and P. F. Mantica and B. A. Brown and R. Broda and P. Bhattacharyya and M. P. Carpenter and M. Cinausero and P. J. Daly and A. D. Davies and T. Glasmacher and Z. W. Grabowski and D. E. Groh and M. Honma and F. G. Kondev and W. Królas and T. Lauritsen and S. N. Liddick and S. Lunardi and N. Marginean and T. Mizusaki and D. J. Morrissey and A. C. Morton and W. F. Mueller and T. Otsuka and T. Pawlat and D. Seweryniak and H. Schatz and A. Stolz and S. L. Tabor and C. A. Ur and G. Viesti and I. Wiedenhöver and J. Wrzesiński},
   doi = {10.1016/S0370-2693(02)02682-5},
   issn = {03702693},
   issue = {1-2},
   journal = {Physics Letters B},
   month = {10},
   pages = {55-62},
   title = {Structure of $^\{52,54\}$Ti and shell closures in neutron-rich nuclei above $^\{48\}$Ca},
   volume = {546},
   year = {2002}
}

@article{Liddick2004,
   author = {S. N. Liddick and P. F. Mantica and R. V. F. Janssens and R. Broda and B. A. Brown and M. P. Carpenter and B. Fornal and M. Honma and T. Mizusaki and A. C. Morton and W. F. Mueller and T. Otsuka and J. Pavan and A. Stolz and S. L. Tabor and B. E. Tomlin and M. Wiedeking},
   doi = {10.1103/PhysRevLett.92.072502},
   issn = {0031-9007},
   issue = {7},
   journal = {Physical Review Letters},
   month = {2},
   pages = {072502},
   title = {Lowest Excitations in $^\{56\}$Ti and the Predicted N=34 Shell Closure},
   volume = {92},
   year = {2004}
}

@article{Liddick2004_2,
   author = {S. N. Liddick and P. F. Mantica and R. Broda and B. A. Brown and M. P. Carpenter and A. D. Davies and B. Fornal and T. Glasmacher and D. E. Groh and M. Honma and M. Horoi and R. V. F. Janssens and T. Mizusaki and D. J. Morrissey and A. C. Morton and W. F. Mueller and T. Otsuka and J. Pavan and H. Schatz and A. Stolz and S. L. Tabor and B. E. Tomlin and M. Wiedeking},
   doi = {10.1103/PhysRevC.70.064303},
   issn = {0556-2813},
   issue = {6},
   journal = {Physical Review C},
   month = {12},
   pages = {064303},
   title = {Development of shell closures at N=32,34. I. $\beta$ decay of neutron-rich Sc isotopes},
   volume = {70},
   year = {2004}
}

@article{Fornal2004,
   author = {B. Fornal and S. Zhu and R. V. F. Janssens and M. Honma and R. Broda and P. F. Mantica and B. A. Brown and M. P. Carpenter and P. J. Daly and S. J. Freeman and Z. W. Grabowski and N. J. Hammond and F. G. Kondev and W. Królas and T. Lauritsen and S. N. Liddick and C. J. Lister and E. F. Moore and T. Otsuka and T. Pawłat and D. Seweryniak and B. E. Tomlin and J. Wrzesiński},
   doi = {10.1103/PhysRevC.70.064304},
   issn = {0556-2813},
   issue = {6},
   journal = {Physical Review C},
   month = {12},
   pages = {064304},
   title = {Development of shell closures at ��=32,34. II. Lowest yrast excitations in even-even Ti isotopes from deep-inelastic heavy-ion collisions},
   volume = {70},
   year = {2004}
}

@article{Prisciandaro2001,
   author = {J. I. Prisciandaro and P. F. Mantica and B. A. Brown and D. W. Anthony and M. W. Cooper and A. Garcia and D. E. Groh and A. Komives and W. Kumarasiri and P. A. Lofy and A. M. Oros-Peusquens and S. L. Tabor and M. Wiedeking},
   doi = {10.1016/S0370-2693(01)00565-2},
   issn = {03702693},
   issue = {1-4},
   journal = {Physics Letters B},
   month = {6},
   pages = {17-23},
   title = {New evidence for a subshell gap at N=32},
   volume = {510},
   year = {2001}
}

@article{Burger2005,
   author = {A. Bürger and T.R. Saito and H. Grawe and H. Hübel and P. Reiter and J. Gerl and M. Górska and H.J. Wollersheim and A. Al-Khatib and A. Banu and T. Beck and F. Becker and P. Bednarczyk and G. Benzoni and A. Bracco and S. Brambilla and P. Bringel and F. Camera and E. Clément and P. Doornenbal and H. Geissel and A. Görgen and J. Grębosz and G. Hammond and M. Hellström and M. Honma and M. Kavatsyuk and O. Kavatsyuk and M. Kmiecik and I. Kojouharov and W. Korten and N. Kurz and R. Lozeva and A. Maj and S. Mandal and B. Million and S. Muralithar and A. Neußer and F. Nowacki and T. Otsuka and Zs. Podolyák and N. Saito and A.K. Singh and H. Weick and C. Wheldon and O. Wieland and M. Winkler},
   doi = {10.1016/j.physletb.2005.07.004},
   issn = {03702693},
   issue = {1-2},
   journal = {Physics Letters B},
   month = {8},
   pages = {29-34},
   title = {Relativistic Coulomb excitation of neutron-rich $^\{54,56,58\}$Cr: On the pathway of magicity from N=40 to N=32},
   volume = {622},
   year = {2005}
}

@article{Al-Khalili2000,
   author = {Jim Al-Khalili},
   doi = {10.1088/2058-7058/13/8/26},
   issn = {0953-8585},
   issue = {8},
   journal = {Physics World},
   month = {8},
   pages = {24-25},
   title = {Nuclear magic numbers appear and disappear},
   volume = {13},
   year = {2000}
}

@article{Karthika2021,
   author = {C. Karthika and C. Kokila and M. Balasubramaniam},
   doi = {10.15415/jnp.2021.91018},
   issn = {2321-9289},
   issue = {1},
   journal = {Journal of Nuclear Physics, Material Sciences, Radiation and Applications},
   month = {8},
   pages = {109-115},
   title = {Appearance / Disappearance of Magic Number in Light Nuclei},
   volume = {9},
   year = {2021}
}

@article{Otsuka2001,
   author = {Takaharu Otsuka and Rintaro Fujimoto and Yutaka Utsuno and B. Alex Brown and Michio Honma and Takahiro Mizusaki},
   doi = {10.1103/PhysRevLett.87.082502},
   issn = {0031-9007},
   issue = {8},
   journal = {Physical Review Letters},
   month = {8},
   pages = {082502},
   title = {Magic Numbers in Exotic Nuclei and Spin-Isospin Properties of the <math display="inline"> <mi mathvariant="italic">NN</mi> </math> Interaction},
   volume = {87},
   year = {2001}
}

@article{Schwengner2020,
   author = {R. Schwengner and R. Massarczyk and R. Beyer and M. Bhike and B. A. Brown and Krishichayan and K. Sieja and W. Tornow and D. Bemmerer and M. Butterling and V. Derya and M. Dietz and F. Fiedler and U. Friman-Gayer and A. Frotscher and M. Grieger and A. Hartmann and A. R. Junghans and T. Kögler and F. Ludwig and B. Lutz and H. Pai and T. Szücs and M. P. Takács and A. Wagner},
   doi = {10.1103/PhysRevC.101.064303},
   issn = {2469-9985},
   issue = {6},
   journal = {Physical Review C},
   month = {6},
   pages = {064303},
   title = {Electric and magnetic dipole strength in $^\{54\}$Fe},
   volume = {101},
   year = {2020}
}

@article{Wilhelmy2018,
   author = {J. Wilhelmy and B. A. Brown and P. Erbacher and U. Gayer and J. Isaak and Krishichayan and B. Löher and M. Müscher and H. Pai and N. Pietralla and P. Ries and D. Savran and P. Scholz and M. Spieker and W. Tornow and V. Werner and A. Zilges},
   doi = {10.1103/PhysRevC.98.034315},
   issn = {2469-9985},
   issue = {3},
   journal = {Physical Review C},
   month = {9},
   pages = {034315},
   title = {Investigation of  ��=1 states and their ��-decay behavior in $^\{52\}$Cr},
   volume = {98},
   year = {2018}
}

@article{Shizuma2017,
   author = {T. Shizuma and T. Hayakawa and I. Daito and H. Ohgaki and S. Miyamoto and F. Minato},
   doi = {10.1103/PhysRevC.96.044316},
   issn = {2469-9985},
   issue = {4},
   journal = {Physical Review C},
   month = {10},
   pages = {044316},
   title = {Low-lying dipole strength in $^\{52\}$Cr},
   volume = {96},
   year = {2017}
}

@article{Krishichayan2015,
   author = {Krishichayan and Megha Bhike and W. Tornow and G. Rusev and A. P. Tonchev and N. Tsoneva and H. Lenske},
   doi = {10.1103/PhysRevC.91.044328},
   issn = {0556-2813},
   issue = {4},
   journal = {Physical Review C},
   month = {4},
   pages = {044328},
   title = {Polarized photon scattering off $^\{52\}$Cr: Determining the parity of $J=1$ states},
   volume = {91},
   year = {2015}
}
